\documentclass[12pt]{article}
\usepackage{graphicx}
\usepackage{amsfonts}
\usepackage{amssymb,amsmath,amscd}
\usepackage{amssymb}
\usepackage[top=2.75cm, bottom=2.75cm, left=2.75cm, right=2.75cm]{geometry}

\newtheorem{proposition}{Proposition}

\newtheorem{remark}[proposition]{Remark}
\newtheorem{definition}[proposition]{Definition}
\newtheorem{example}[proposition]{Example}

\def\+{{+\!\!\!+}}

\def\d{\partial}

\def\N{\nabla}

\def\pmb#1{\setbox0=\hbox{#1}%
\kern.0em\copy0\kern-\wd0
\kern-.04em\copy0\kern-\wd0
\kern.08em\copy0\kern-\wd0
\kern-.04em\raise.0433em\box0 }         

\newcommand{\nc}{\newcommand}
\nc{\beq}{\begin{equation}}
\nc{\eeq}[1]{\label{#1}\end{equation}}
\nc{\ber}{\begin{eqnarray}}
\nc{\eer}[1]{\label{#1}\end{eqnarray}}
\nc{\pek}[1]{\cite{#1}}
\nc{\enr}[1]{(\ref{#1})}
\nc{\kal}[1]{{\cal{#1}}}
\nc{\dott}{\;\cdot\;}
\nc{\coker}{\mathrm{coker}}
\nc{\ie}{{\it i.e.}}
\nc{\eg}{{\it e.g.}}
\newcommand{\Section}[1]{\section{#1} \setcounter{equation}{0}}

\def\R{{\mathbb R}}
\def\S{{\mathbb S}}
\def\superx{{\mathsf x}}
\def\supery{{\mathsf y}}
\def\supereta{{\mathsf e}}
\def\superX{{\mathsf X}}
\def\superEta{\boldsymbol{\eta}}

\def\Sigmagn{{\Sigma_{g,n}}}
\def\Sigmag{{\Sigma_g}}
\def\Sigmaguno{{\Sigma_{g,1}}}
\def\0 {\nonumber}
\def\M{{\cal M}}
\def\N{{\cal N}}
\def\abar{{\bar a}}
\begin{document}

\setcounter{page}{0}
\newcommand{\inv}[1]{{#1}^{-1}} 
\renewcommand{\theequation}{\thesection.\arabic{equation}}
\newcommand{\dd}{\partial}
\newcommand{\ra}{\rightarrow}
\newcommand{\id}{\mathrm{id}}
\newcommand{\be}{\begin{equation}}
\newcommand{\ee}{\end{equation}}
\newcommand{\C}{\mathcal{C}}
\newcommand{\bt}{\bullet}
\newcommand{\FF}{\mathcal{F}}
\newcommand{\gh}{\mathrm{gh}}
\newcommand{\1}{{\bf 1}}
\newcommand{\ZZ}{\mathbb{Z}}
\newcommand{\Fun}{\mathrm{Fun}}
\newcommand{\PP}{\mathcal{P}}
\newcommand{\Maps}{\mathrm{Maps}}
\newcommand{\fib}{\mathrm{fib}}
\newcommand{\ev}{\mathrm{ev}}
\newcommand{\g}{\mathfrak{g}}
\newcommand{\RR}{\mathbb{R}}
\newcommand{\W}{\mathcal{W}}
\newcommand{\Tr}{\mathrm{Tr}}
\newcommand{\Hom}{\mathrm{Hom}}
\newcommand{\J}{\mathcal{J}}
\newcommand{\bea}{\begin{eqnarray}}
\newcommand{\eea}{\end{eqnarray}}
\newcommand{\re}[1]{(\ref{#1})}
\newcommand{\qv}{\quad ,}
\newcommand{\qp}{\quad .}
\def\O{{\cal O}}

\thispagestyle{empty}
\begin{flushright} \small
UUITP-27/08  \\
NORDITA-2008-65 \\
\end{flushright}
\smallskip
\begin{center} \LARGE
{\bf Finite dimensional AKSZ-BV theories}
 \\[12mm] \normalsize
{\large\bf Francesco Bonechi$^{a}$, Pavel Mn\"ev$^b$ and Maxim Zabzine$^{c}$} \\[8mm]
 {\small\it
$^a$I.N.F.N. and Dipartimento di Fisica,\\
  Via G. Sansone 1, 50019 Sesto Fiorentino - Firenze, Italy \\
~\\
$^b$Petersburg Department of V. A. Steklov Institute of Mathematics,\\
  Fontanka 27,  191023 St. Petersburg, Russia \\
~\\
$^c$Department of Physics and Astronomy,
Uppsala University, \\ Box 516, SE-751 20 Uppsala, Sweden
\\~\\
}
\end{center}
\vspace{10mm}
\centerline{\bfseries Abstract} \bigskip
We describe a canonical reduction of AKSZ-BV theories to the cohomology
 of the source manifold. We get a finite dimensional BV theory that describes
 the contribution of the zero modes to the full QFT. Integration can be defined
 and correlators can be computed.  As an illustration of the general construction
   we consider two dimensional Poisson sigma model and three dimensional Courant
    sigma model. When the source manifold is compact,
  the reduced theory is a generalization of the AKSZ construction where we
  take as source the cohomology ring. We present the possible generalizations of the AKSZ theory.

\noindent

\eject
\normalsize



\section{Introduction}
\label{start}

The Batalin-Vylkovisky (BV) formalism \cite{Batalin:1981jr} is widely regarded as the most powerful and general approach to
 the quantization of gauge theories.  The idea is to extend the space of fields of the original gauge theory by
  auxiliary fields (ghosts, antighosts, Lagrangian multipliers etc.) and their  conjugate antifields in such way that
   the total field-antifield space is equipped with two canonical structures: an odd Poisson bracket (antibracket)
    and an odd second order differential operator $\Delta$ (BV Laplacian).  The original gauge action should
     be extended to master action defined on the total field-antifield space in such way that the
       master action satisfies the master equation (the equation involving  antibracket
        and  BV Laplacian).   The path  integral is defined as an appropriate integral over half of the field-antifield space.
         The master action can be computed within the homological perturbation theory which can be very difficult to carry out in general.   In \cite{Alexandrov:1995kv} Alexandrov, Kontsevich, Schwarz and Zaboronsky (AKSZ) proposed
  a way to construct solutions of the BV classical
master equation directly, without  any reference to a classical action with
a set of gauge symmetries.   Their approach (AKSZ method) uses mapping spaces of supermanifolds equipped with the
  additional structures.
The AKSZ method was applied in \cite{Alexandrov:1995kv} to
the Chern-Simons theory, the Witten A- and
B-models. Furthermore the AKSZ approach was applied to two-dimensional Poisson sigma model in \cite{Cattaneo:2001ys}
 and to  three-dimensional Courant sigma model in \cite{Ikeda:2002wh}, \cite{Roytenberg:2006qz}.
  Moreover the higher dimensional case of open p-branes were discussed
   in \cite{Park:2000au, Hofman:2002rv, Hofman:2002jz}.

 In the present work we propose the reduction within the AKSZ framework to a finite dimensional BV theory
  which governs the zero mode contribution and is responsible for the semiclassical approximation in the full theory.
   These finite dimensional BV theories offer interesting perspective on some of the standard geometry.
   The proposed reduction naturally suggests a generalization of AKSZ framework.

The paper is organized as follows.  In Section \ref{BVsummary} we sketch the essentials of the BV formalism.
Section \ref{AKSZgeneral} reviews the AKSZ formalism and discusses its reduction in general terms. The rest
 of the paper is devoted to the consideration of the examples of this reduction and to the discussion of
  finite dimensional BV theories arising as a result of the reduction.  Section \ref{PSM} treats the Poisson
   sigma model defined over a closed Riemann surface of genus $g$.   Section \ref{CSM} considers the
   Courant sigma model, a three dimensional topological sigma model.
Section \ref{boundary} considers the more involved setup of the Poisson sigma model on the Riemann surface
 with a boundary.
 In Section \ref{generalizationAKSZ} we discuss the generalization of the AKSZ construction. In Section \ref{summary} we give a summary and discuss the possible developments.
  In Appendix \ref{ap-ber} we calculate the Berezinian measure for the reduced BV manifold of the Courant
   sigma model. In Appendix \ref{relative_homology} we collect some facts on the relative cohomology of the manifolds with boundaries.

\section{Summary of BV formalism}
\label{BVsummary}

In this Section we recall the basic notions of BV formalism and fix the notation.
 For further details the reader may consult the following
   reviews \cite{cattaneobv, fiorenza}.

    The BV algebra can be defined in many different but equivalent ways. In particular,
 a Gerstenhaber algebra (odd Poisson algebra) $({\cal A}, \{~,~\})$ together with an odd $\mathbb{R}$--linear map
 $$\Delta ~:~{\cal A} \longrightarrow {\cal A}\;,$$
  which squares to zero $\Delta^2=0$ and generates the bracket $\{~,~\}$ as
   $$\{f, g\}=  (-1)^{|f|} \Delta(fg)+(-1)^{|f|+1}(\Delta f)g - f(\Delta g) \;,$$
  is called a BV-algebra. $\Delta$ is called odd Laplace operator (odd Laplacian).
   Quite often such odd Poisson bracket is called  antibracket.

The canonical example of BV algebra is given by the space of functions on $W\oplus \Pi W^*$,
where $W$ is a superspace, $W^*$ is its dual and $\Pi$ stands for the reversed parity functor.
$W\oplus \Pi W^*$ is equipped with an odd non-degenerate pairing. Let $y^a$ be the coordinates
on $W$ (the fields) and $y_a^+$ be the corresponding coordinates
on $\Pi W^*$ (the antifields). We denote the parity of $y^a$ as $(-1)^{|y^a|}$ and that of $y^+_a$ as $(-1)^{|y^+_a|}=(-1)^{|y^a|+1}$. Then the odd Laplacian is defined as follows
     \beq
      \Delta = (-1)^{|y_a|}\frac{\d}{\d y_a^+}\frac{\d}{\d y^a}~.
     \eeq{trivialoddlap}
 It generates the canonical antibracket on $C^\infty ( W\oplus \Pi W^*)$
 \beq
  \{ f, g\} =
   (-1)^{|y^a|}\frac{\overleftarrow{\d}f}{\d y_a^+} \frac{\overrightarrow{\d}g}{\d y^a} +
   (-1)^{|y^a|}\frac{\overleftarrow{\d}f}{\d y^a} \frac{\overrightarrow{\d}g}{\d y_a^+}
   ~,
 \eeq{generalcxlsoo}
  where we use the notation $\overrightarrow{\d}_v f = \d_v f$ and $\overleftarrow{\d}_v f = (-1)^{|v||f|}
   \d_v f$.  Indeed the bracket (\ref{generalcxlsoo}) is non degenerate and defines the canonical odd symplectic structure on $W\oplus \Pi W^*$.

   A Lagrangian submanifold ${\cal L} \subset W\oplus \Pi W^*$ is an isotropic supermanifold of maximal
    dimension. The volume form $dy^1...dy^n dy_1^+...dy_n^+$ induces a well defined volume form
     on ${\cal L}$. Thus the integral
     \beq
       \int\limits_{\cal L} f,~~~~~~~f\in C^\infty(W\oplus \Pi W^*)
     \eeq{definelagen}
 is defined for any ${\cal L}$.   If $\Delta f =0$, then then the integral (\ref{definelagen}) depends only on the
 homology class of ${\cal L}$. Moreover  the integral (\ref{definelagen}) is zero for any Lagrangian ${\cal L}$
    if $f= \Delta g$.

 The canonical example $W\oplus \Pi W^*$ can be generalized to the
cotangent bundle $T^*[-1] {\cal M}$ of any graded manifold ${\cal M}$ \cite{Schwarz:1992nx}.
  By a graded manifold we mean a sheaf of freely generated $\mathbb{Z}$-graded commutative algebras
    over a smooth manifold \cite{Voronov:2001qf}.
As a cotangent bundle, $T^*[-1] {\cal M}$ is naturally equipped with an odd Poisson bracket that makes
$C^\infty (T^*[-1] {\cal M})$ to a Gerstenhaber algebra.
The idea is that locally one can map $T^*[-1] {\cal M}$ to $W\oplus \Pi W^*$, define the bracket on coordinates with (\ref{generalcxlsoo}) and then glue the patches in a consistent manner.

 Now in order to  define the odd Laplacian $\Delta$ we need an integration over $T^*[-1] {\cal M}$.
 Namely, the choice of a volume form $v$ on ${\cal M}$ produces the corresponding volume\footnote{We require
  that the measure is well-defined functional on the space of compactly supported functions and locally
   it is Berezinian measure.}
  form $\mu_v$ on $T^*[-1]{\cal M}$. The divergence operator is defined as a map
     from the vector fields on $T^*[-1]{\cal M}$ to $C^\infty(T^*[-1]{\cal M})$ through the
    following  integral relation
   \beq
  \int\limits_{T^*[-1]{\cal M}}   X(f) ~\mu_v =   - \int\limits_{T^*[-1]{\cal M}}  {\rm div}_{\mu_v} X ~ f ~\mu_v~,~~~~~~~~
  \forall f \in C^\infty (T^*[-1] {\cal M})~,
 \eeq{divergencedef}
 with $X$ being a vector field. As one can easily check, for any function $f$ and vector field $X$ the divergence satisfies
\begin{equation}\label{divergence}
 {\rm div}_{\mu_v}(f X)= f {\rm div}_{\mu_v}(X) + (-1)^{|f||X|} X(f)\;.
\end{equation}

Now the odd Laplacian of $f\in C^\infty(T^*[-1]\cal M)$ is defined through the divergence of the corresponding Hamiltonian vector field as
  \beq
   \Delta_v f = \frac{(-1)^{|f|}}{2}{\rm div_{\mu_v}} X_f~,~~~~~\{f, g\} =X_f(g)~.
  \eeq{definlapppekdk}
 Indeed one can check that thanks to (\ref{divergence}) $\Delta_v$ generates the bracket and $\Delta_v^2=0$.  Thus
  $C^\infty (T^*[-1] {\cal M})$ is a BV-algebra as defined above, see \cite{kosmann3}
   for the explicit calculations. If the volume form is written in terms of an even density $\rho_v$ as
   $$\mu_v=\rho_v dy^1\cdots dy^n dy^+_1\cdots dy^+_n~,$$
   then the Laplacian can be written as
  \begin{equation}\label{global_laplacian}
  \Delta_v= (-1)^{|y_a|}\frac{\d}{\d y_a^+}\frac{\d}{\d y^a} + \frac{1}{2} \{\log\rho_v, -\}~.
  \end{equation}
   This Laplacian squares to zero since we take a specific Berezinian measure $\rho_v$ which
    originates from the measure on $\M$. Thus $T^*[-1]\cal M$ is equipped with the SP-structure in
     the  Schwarz's terminology \cite{Schwarz:1992nx} (also see \cite{Khudaverdian:1995np} for a nice review of  the related issues)
      and in what follows we refer to them simply as BV-manifolds.

 There exists a canonical way (up to a sign) of restricting a volume form $\mu_v$ on
  $T^*[-1] {\cal M}$ to a volume form on a Lagrangian submanifold ${\cal L}$. We denote
   such restriction as $\sqrt{\mu_v}$ and consider  the integrals of the form
  \beq
   \int\limits_{\cal L} \sqrt{\mu_v}~ f~,~~~~~~~~ f \in C^\infty (T^*[-1] {\cal M})~.
  \eeq{integhajw28920}
 Again if $f$ is $\Delta$-closed then the integral depends only on homology class of ${\cal L}$
  and if $f$ is $\Delta$-exact then the integral is zero.
 In particular we are interested in the situation when the integrands in (\ref{integhajw28920}) are of the form
 \beq
 \int\limits_{\cal L}  \sqrt{\mu_v} ~\Psi e^{S} \equiv \langle \Psi \rangle~,
\eeq{confjdjdkkkll}
 where we assume naturally that $\Delta_v (\Psi e^{S})=0$.
If $\Psi=1$ then we get the following relation
\beq
 \Delta_v \left ( e^S \right ) =0~~ \Longleftrightarrow~~ \Delta_v S + \frac{1}{2} \{S, S\}=0~,
\eeq{rela327388}
 which is known as the {\it quantum master equation}. In the general case we have
\beq
\Delta_v \left ( \Psi e^S \right ) =0~~ \Longleftrightarrow~~   \Delta_{(v,S)} \Psi = \Delta_v \Psi + \{ S , \Psi\}=0~,
\eeq{definitionquantum}
 where we refer to $\Delta_{(v,S)}$ as the quantum Laplacian.  In the derivation
  of (\ref{definitionquantum}) we have used the quantum master equation (\ref{rela327388}). A function
   $S$ that satisfies the quantum master equation is called a quantum  BV action and $\Psi$ satisfying (\ref{definitionquantum}) is a quantum observable. Indeed the quantum observables are elements of the
cohomology $H(\Delta_{(v,S)})$; by the above construction it is clear that $S$ defines the isomorphism
  \beq
   H^\bullet (\Delta_v) \approx H^\bullet(\Delta_{(v,S)})\;.
  \eeq{isomrocohomls;}
   The integral (\ref{confjdjdkkkll}) defines a trace ($H^\bullet(\Delta_{(v,S)}) \rightarrow \mathbb{C}$) on this cohomology.

 If we change $S$ to $S/\hbar$, we see that in the classical limit ($\hbar \rightarrow 0$) $S$ must satisfy the classical master equation $\{ S, S\}=0$ and the classical observables $\Psi$ are such that $\delta_{BV} \Psi \equiv \{ S, \Psi \}=0$. Due to the classical master equation the vector field $\delta_{BV}$ squares to zero and defines the cohomology $H(\delta_{BV})$ of classical observables.

 If ${\cal M}$ is a finite dimensional manifold then everything is well-defined. In the following Sections we provide several finite dimensional BV manifolds equipped with a solution of the classical (quantum) master equation.
 However in field theory
  one deals with ${\cal M}$ being infinite dimensional.
 In fact, $\cal M$ is usually the space of the physical fields, ghosts and Lagrange multipliers, that is infinite dimensional. This set of fields is then extended by adding antifields such that together they form $T^*[-1]{\cal M}$, where an odd Poisson bracket is well-defined on large enough class of functions, as described above.
 However there is no well-defined measure on ${\cal M}$ and thus there is no well-defined odd Laplace operators. In physics literature, the naive Laplacian of the form  (\ref{generalcxlsoo}) is used.
 Moreover the field theory suffers from the problems with renormalization which can be resolved within the perturbative setup.

\bigskip
\bigskip

\section{AKSZ formalism and its reduction}
\label{AKSZgeneral}

In this Section we review the AKSZ construction \cite{Alexandrov:1995kv} of the solutions of the classical master equation.
 Here we will follow the presentation given in \cite{Roytenberg:2006qz} and we use the language
  of graded manifolds. Relevant definitions are postponed to Section \ref{generalizationAKSZ}. For further details the reader may consult \cite{Voronov:2001qf}.

The AKSZ solution of the classical master equation is defined starting from the following data:

\medskip
\noindent{\bf The source}: A graded manifold $\N$ endowed with a cohomological vector field $D$ and
a measure $\int\limits_\N\mu$ of degree $-n-1$ for some positive integer $n$.
 In what follows the source will be $\N=T[1]\N_0$, for any smooth manifold $\N_0$ of dimension $n+1$, with
$D=d$ the de Rham differential over $\N_0$ and the canonical coordinate measure.

\noindent{\bf The target}: A graded symplectic manifold $(\M,\omega)$ with $\deg(\omega)=n$ and an
homological vector field $Q$ preserving $\omega$.  We require that $Q$ is Hamiltonian,
{\it i.e.} it exists $\Theta\in C_{n+1}(\M)$ (functions of degree $n+1$) such that $Q=\{\Theta,-\}$.  Therefore $\Theta$ satisfies
 the following Maurer-Cartan equation
 $$\{ \Theta, \Theta \}=0~.$$

Since graded manifolds are ringed spaces, we consider the space ${\rm Maps}(\N,\M)$ to be the space of morphisms of ringed spaces (see Definition \ref{morph_ring_space}). We choose a set of coordinates $X^A=\{x^\mu;\xi^m\}$ on the target $\M$, where $\{x^\mu\}$ are the coordinates for an open $U\subset \M_0$ and $\{\xi^m\}$ are the coordinates in the formal directions.  We also choose the coordinates $\{u^\alpha;\theta^a\}$ on the source ${\cal N}$ correspondently. Let $(\phi,\Phi)\in{\rm Maps}(\N,\M)$, where $\phi:\N_0\rightarrow\M_0$. The superfield $\Phi$ is defined as an expansion over the formal coordinates
  of $\N$ for $\phi^{-1}(U)$
\begin{equation}\label{superfield}
\Phi^A= \Phi_0^A(u) + \Phi_{a}^A(u) \theta^a + \Phi_{a_1a_2}^A(u) \theta^{a_1}\theta^{a_2} + \ldots ~,
\end{equation}
such that $\Phi^\mu_0 = x^\mu\circ\phi$. The symplectic form $\omega$ of degree
 $n$ on $\M$ can be written in the Darboux coordinates $\omega = dX^A \omega_{AB} dX^B$.
 Using this form we define the symplectic form of degree $-1$ on ${\rm Maps}(\N,\M)$  as
\begin{equation}
 \label{P_structure}
 \Omega = \int\limits_\N \mu ~~\delta \Phi^A ~\omega_{AB}~ \delta \Phi^B~.
\end{equation}
 Thus the space of maps ${\rm Maps}(\N,\M)$ is naturally equipped with the odd Poisson bracket $\{~,~\}$.
 Since the space ${\rm Maps}(\N,\M)$ is infinite dimensional we cannot define the BV Laplacian properly. We can only talk about the naive Laplacian adapted to the local field-antifield splitting, given  by the formula (\ref{trivialoddlap}).
 However on ${\rm Maps}(\N,\M)$ we can discuss the solutions of the classical master equation. The AKSZ action then reads
\beq
S_{BV}[\Phi]= S_{kin}[\Phi]+ S_{int}[\Phi] = \int\limits_\N \mu ~\left ( \frac{1}{2} \Phi^A\omega_{AB} D\Phi^B +
(-1)^{n+1} \Phi^*(\Theta)\right )~.
\eeq{AKSZ-action}
and solves the classical master equation $\{S_{BV},S_{BV}\}=0$ with respect to the bracket
 defined by the symplectic structure (\ref{P_structure}). We denote with $\delta_{BV}$
  the Hamiltonian vector field for $S_{BV}$, which is homological as a consequence of classical
   master equation. The action (\ref{AKSZ-action}) is invariant under all orientation preserving diffeomorphisms of $\N_0$
    and thus defines a topological field theory.  The solutions of the classical field equations of (\ref{AKSZ-action}) are graded differentiable maps $(\N, D) \rightarrow (\M, Q)$, i.e. maps which commute with the homological vector fields.

The homological vector field $Q$ on $\M$ defines a complex on $C^\infty(\M)$ whose
cohomology we denote with $H_Q(\M)$. Take $f\in C^\infty(\M)$ and expand $\Phi^*f$
in the formal variables on $\N$, {\it i.e.}
$$\Phi^*f= O^{(0)}(f) + O^{(1)}_a(f) \theta^a + O^{(2)}_{a_1 a_2}(f) \theta^{a_1} \theta^{a_2} + \ldots ~.$$
We compute
$$
\delta_{BV}(\Phi^*f)=\{S_{BV} ,\Phi^*f\} = D\Phi^*f + \Phi^* Qf~,
$$
so that, if $Qf=0$ and $\mu_k$ is a $D$-invariant linear functional on $C_k(\N)$, {\it i.e.}
a representative of a homology class of $\N_0$, then $\mu_k(O^{(k)}(f))$ is
$\delta_{BV}$-closed, {\it i.e.} it is a classical observable. Therefore $H_Q(\M)$ naturally defines
a set of classical observables in the theory.

Like in the smooth case, the symplectic reduction of the odd symplectic manifold ${\rm Maps}(\N,\M)$ is  specified by the choice of a coisotropic submanifold. The reduced space naturally inherits the odd symplectic structure and any function on ${\rm Maps}(\N,\M)$ descends to the quotient
provided it is invariant once restricted to the coisotropic submanifold.
 In this paper we consider the symplectic reduction of ${\rm Maps}(\N,\M)$ to the
cohomology $H_D(\N)=H_{dR}(\N_0)$ of the source. We consider
the constraint
$$\Lambda= \int\limits_{\cal N}\mu~\Lambda_A D\Phi^A~,$$
  which defines the coisotropic submanifold $D\Phi = 0$ of ${\rm Maps}(\N,\M)$ (the constrained surface).
  The corresponding infinitesimal  gauge transformations
\begin{equation}\label{gauge_constant_config}
\delta_\Lambda \Phi^A = D \Lambda^A
\end{equation}
identify two configurations which differ by a $D$-exact term. If $\N_0$
 has a boundary then boundary conditions must be discussed. We leave
 the case with boundary to  Section \ref{boundary} and for now we assume that $\N_0$ does
  not have boundary. The reduced odd symplectic manifold is finite dimensional and can be globally described as the space of maps from the cohomology of $\N_0$ to $\M$. More precisely, we can consider
a generalized AKSZ construction where the source is the cohomology of $\N_0$ seen as a sheaf $X_{\N_0}$ of graded commutative algebras over a point, equipped with the zero homological vector field and the integration naturally induced on cohomology. The source is not anymore a graded manifold since the cohomology ring is not freely generated in general. However the space of maps ${\rm Maps}(X_{\N_0},\M)$
is still defined; this simply corresponds to interpret  $\{\theta^a\}$ in
the superfield (\ref{superfield}) as the generators of the cohomology ring $H_{dR}(\N_0)$. The simplest case is when $\N_0$ has cohomology ring concentrated in degree zero and top degree.
    In this case the reduced theory will be simply $T^*[-1]\M$.

For any $f\in C^\infty(\M)$, $\Phi^*f-(\Phi+D\Lambda)^*f$ is $D$-exact. Thus
any observable $\mu_k(O^{(k)}(f))$ in the full theory is invariant under the gauge transformations
 (\ref{gauge_constant_config}) and defines a function on the reduced odd symplectic manifold.
  In particular, the BV-action (\ref{AKSZ-action}) $S_{BV}=\int\limits_\N \mu ~\Phi^*(\Theta)$
  defines a solution of the classical master equation that coincides
  with the AKSZ action on ${\rm Maps}(X_{\N_0},\M)$.

It is crucial that the reduced odd symplectic manifold is finite dimensional so that integration and BV Laplacian can be well defined. Thus we can define properly the quantum master equation and discuss the possible obstructions to satisfy it. The rest of the paper consists of a detailed
account of these reduced BV-theories in several examples.

\section{Two dimensional case: Poisson sigma model}
\label{PSM}

Let us consider the AKSZ construction described in the previous section with
source $\N=T[1]\Sigmag$,  where $\Sigmag$ is the two dimensional compact
surface of genus $g$. Then the target $\M$ is a symplectic graded manifold
of degree $1$ with a homological vector field $Q$ preserving the symplectic
structure. It must be necessarily of the form $\M=T^*[1]M$ where $M$ is a Poisson
manifold with Poisson tensor $\alpha$. By choosing coordinates $\{x^\mu,\beta_\mu\}$,
 the Hamiltonian for $Q$ is $\Theta=\alpha^{\mu\nu}\beta_\mu\beta_\nu\in C_2(\M)$.
 We have that $C^\infty(\M)=C^\infty(\Lambda TM)$ and the cohomology $H_Q(M)$
 of the complex $(C^\infty(\M),Q)$ coincides with the usual Lichnerowicz-Poisson
 cohomology $H_{LP}(M;\alpha)$. This is the case studied in \cite{Cattaneo:1999fm}
  as a BV quantization of the Poisson Sigma Model. We will first review the relevant
   results from \cite{Cattaneo:1999fm} and \cite{Cattaneo:2001ys} and then study the
   reduction procedure to the finite
  dimensional $BV$-manifold outilined in the previous section. In the case of
   $\Sigma_0=\S^2$ the reduction to the finite dimensional theory has been
   studied in \cite{Bonechi:2007ar}.

   The Poisson Sigma Model covers many interesting geometrical structures.
 For example,  if $A$ is a Lie algebroid then the dual vector bundle $A^*$ regarded as a manifold is naturally equipped with a Poisson structure.  This is an example of a non-compact Poisson manifold where some issues related to the integration require extra care. We assume that for non-compact case
       the integration is defined as functional over the space of functions with compact support (or with  exponential
        decay along the non-compact directions)  and thus
        formally all our considerations are equally applicable both for compact and non-compact cases.

\subsection{BV action}
\label{BVaction}

Let $\{u^\alpha\}$ be coordinates on $\Sigma_g$ and $\{x^\mu\}$ on $M$. The superfields read as
$$ {\mathsf X}^\mu = X^\mu + \theta^\alpha \eta_\alpha^{+\mu} - \frac{1}{2} \theta^\alpha\theta^\beta
\beta^{+\mu}_{\alpha\beta}~,$$
$$ \boldsymbol{\eta}_\mu = \beta_\mu + \theta^\alpha \eta_{\alpha\mu}
+ \frac{1}{2} \theta^\alpha\theta^\beta X^{+}_{\alpha\beta\mu}~,$$
  with $\{\theta^\alpha\}$ being the odd coordinates on $T[1]\Sigmag$.
In the BV language, the components of ghost number $0$, $X$ and $\eta$
 are the classical fields, $\beta$ is a ghost with the ghost number $1$, while
 $\eta^+$, $\beta^+$ and $X^+$ are antifields of ghost number $-1$, $-2$ and
 $-1$ respectively. The space of maps ${\rm Map}(T[1]\Sigmag,T^*[1]M)$
  can be seen as $T^*[-1]{\cal M}$, where ${\cal M}$ is the infinite dimensional manifold
   corresponding to the fields $(X,\eta,\beta)$.

If we change coordinates $y^i=y^i(x)$, the superfields transform as
\begin{equation}\label{change_coordinates_superfields}
{\mathsf Y}^i = y^i({\mathsf X}) \;,~~~~~~~ {\boldsymbol{\eta}_i}=
\frac{\partial x^\mu}{\partial y_i}({\mathsf X}){\boldsymbol{\eta}_\mu} \;.
\end{equation}
For later utility we add the explicit component content of (\ref{change_coordinates_superfields})
\begin{equation}\label{change_coordinates_components}
\eta^{+i}_\alpha = \frac{\partial y^i}{\partial x^\mu} \eta^{+\mu}_\alpha \;, ~~~
\beta^{+i}_{\alpha\beta} = \frac{\partial y^i}{\partial x^\mu} \beta^{+\mu}_{\alpha\beta}
+ \frac{\partial^2 y^i}{\partial x^\mu\partial x^\nu} \eta^{+\mu}_\alpha\eta^{+\nu}_\beta~,~~~~
 \beta_i=\frac{\partial x^\mu}{\partial y^i}\beta_\mu \;,
\end{equation}
$$
\eta_i = \frac{\partial x^\mu}{\partial y^i}\eta_\mu + \frac{\partial}{\partial x^\nu}
 \frac{\partial x^\mu}{\partial y^i} \eta^{+\nu} \beta_\mu~,~~$$
$$
X^+_{i\alpha\beta} = \frac{\partial x^\mu}{\partial y^i} X^+_{\mu\alpha\beta} -
 2 \frac{\partial}{\partial x^\nu} \frac{\partial x^\mu}{\partial y^i} \eta^{+\nu}_\alpha\eta_{\mu\beta}
  - \frac{\partial}{\partial x^\nu} \frac{\partial x^\mu}{\partial y^i} \beta^{+\nu}_{\alpha\beta}\beta_\mu
  -\frac{\partial^2 \ }{\partial x^\nu\partial x^\rho}\frac{\partial x^\mu}{\partial y^i}
  \eta^{+\nu}_\alpha\eta^{+\rho}_\beta \beta_\mu\;.
$$
  The AKSZ action defined in (\ref{AKSZ-action}) reads
  \beq
   S_{BV} = \int d^2\theta d^2 u\, \left ( \boldsymbol{\eta}_\mu D {\mathsf X}^\mu +
    \frac{1}{2} \alpha^{\mu\nu}({\mathsf X}) \boldsymbol{\eta}_\mu \boldsymbol{\eta}_\nu\right )~,
  \eeq{fullBVsuperfiel}
 where $D= \theta^\alpha \d_\alpha$. The odd symplectic structure is
   \beq
    \omega = \int\limits_{\Sigmag} \left ( \delta X \wedge \delta X^+ + \delta \eta\wedge \delta \eta^+ +
     \delta \beta \wedge \delta \beta^+ \right )~.
   \eeq{oddsymplecticsty}

   The action (\ref{fullBVsuperfiel})
   satisfies both classical and naive quantum master equations \cite{Cattaneo:1999fm}.
 The corresponding BV operator $\delta_{BV}$ acts on the superfields as follows
\ber
\label{EQ123}&&\delta_{BV} {\mathsf X}^\mu = D{\mathsf X}^\mu +
\alpha^{\mu\nu} ({\mathsf X}) \boldsymbol{\eta}_\nu~ ,\\
\label{EQ1234}&& \delta_{BV} \boldsymbol{\eta}_\mu = D\boldsymbol{\eta}_\mu
 + \frac{1}{2} \d_\mu \alpha^{\nu\rho}({\mathsf X}) \boldsymbol{\eta}_\nu \boldsymbol{\eta}_\rho~ .
\eer{BRSTinsuperfield}
 The local and non-local classical observables are labelled by the Lichnerowicz-
Poisson cohomology \cite{Bonechi:2007ar}.
 In components the AKSZ-BV action  (\ref{fullBVsuperfiel}) has the form
$$S_{BV} = \int\limits_{\Sigmag} \eta_\mu \wedge dX^\mu + \frac{1}{2} \alpha^{\mu\nu} (X)
 \eta_\mu \wedge \eta_\nu + X^+_\mu \alpha^{\mu\nu}(X) \beta_\nu - \eta^{+\mu} \wedge \left (d\beta_\mu
  + \d_\mu \alpha^{\rho\nu}(X) \eta_\rho \beta_\nu\right ) - $$
  \beq
  -\frac{1}{2} \beta^{+\mu} \partial_\mu \alpha^{\rho\nu}(X)
  \beta_\rho\beta_\nu - \frac{1}{4} \eta^{+\mu} \wedge \eta^{+\nu}
  \partial_\mu \partial_\nu \alpha^{\rho\sigma} (X)\beta_\rho \beta_\sigma~ .
\eeq{BVfullaction}

\subsection{The reduced BV-AKSZ theory}
\label{reduction_cohomology}

In this subsection we describe the reduction of the BV manifold ${\rm Maps}(T[1]\Sigmag,T^*[1]M)$
  down to "the constant configurations" as described in Section \ref{AKSZgeneral}.
  We  obtain a solution of the classical master equation on the finite dimensional
 BV manifold. Namely we define the reduction  with respect to the following constraints
\begin{equation}
 \label{constraints_finite}
 \Lambda = \int\limits_{T[1]\Sigma} \Lambda_\mu D {\mathsf X}^\mu \,,~~~~~
 T= \int\limits_{T[1]\Sigma} T^\mu D\boldsymbol{\eta}_\mu \;,
\end{equation}
with $\Lambda_\mu(u,\theta)=\Lambda_{\mu}^{(0)}+\Lambda_{\mu\alpha}^{(1)}\theta^\alpha+
\Lambda_{\mu\alpha\beta}^{(2)}\theta^\alpha\theta^\beta$ and $T^\mu(u,\theta)=T^\mu_{(0)}+
T^\mu_{(1)\alpha} \theta^\alpha + T^\mu_{(2)\alpha\beta}\theta^\alpha\theta^\beta$.
We assign $|\Lambda_\mu|=0$ and $|T^\mu|=-1$. After a short computation one gets
\begin{equation}\label{gauge_transf_superfield}
\delta_{\Lambda,T }{\mathsf X}^\mu=\{T,{\mathsf X}^\mu\} = D T^\mu \,,\;~~~~~~
\delta_{\Lambda,T}\boldsymbol{\eta}_\nu=\{\Lambda,\boldsymbol{\eta}_\nu \}= D \Lambda_\mu \; ,
\end{equation}
or in components
\begin{eqnarray}\label{gauge_transf}
\delta_{\Lambda,T}X^\mu=0\,,~~~~ \delta_{\Lambda,T}\eta^{+\mu} = dT_{(0)}^\mu\;, ~~~~~
\delta_{\Lambda,T} \beta^{+\mu} = d T^\mu_{(1)} \cr
\delta_{\Lambda,T}\beta_\mu= 0 \, ,~~~~~~~ \delta_{\Lambda,T}\eta_\mu =
d\Lambda^{(0)}_\mu\;,~~~~~~~ \delta_{\Lambda,T}X^+_\mu = d\Lambda^{(1)}_\mu\;.
\end{eqnarray}

We see that the reduction with respect to the constraints $D{\mathsf X}=D\boldsymbol{\eta}=0$
 consists in identifying $({\mathsf X},\boldsymbol{\eta})$ with $({\mathsf X}+DT,\boldsymbol{\eta}+D\Lambda)$
  and then in going to the cohomology of $\Sigmag$.

In order to define the reduced variables, let us choose a basis for the homology
$H_\bullet(\Sigmag)$. Let $u_0\in \Sigmag$ be a generator in degree zero and
 the whole surface $\Sigmag$ be the generator in degree two and consider the
  canonical basis  $\{c_I, c^J\}$ in  $H_1(\Sigmag)$, in degree one,  with
 the following intersection numbers
$$ \#(c_I, c_J) = \#(c^I, c^J)=0~,~~~~~~~\#(c_I, c^J) = - \#(c^J, c_I) = \delta^J_I~.$$
 The Poincar\'e dual basis $\{ e^J, e_I\}$ in $H^1(\Sigmag)$ is defined as
 $$ \int\limits_{c_I} e^J =  \int\limits_{c^J} e_I = \delta_I^J~,~~~~~~~ \int\limits_{c_I} e_J =
 \int\limits_{c^I} e^J =0~,$$
  and satisfies
  $$ \int\limits_{\Sigma_g} e^J \wedge e_I =  \delta^J_I~.$$
Let us introduce the reduced coordinates obtained from the zero form fields
$$
x^\mu = X^\mu(u_0) \;,~~~~    b_\mu = \beta_\mu(u_0)\;,
$$
from the one form fields
   $$ \eta_I = \int\limits_{c_I}\eta ~,~~~~~~~\eta^J =  \int\limits_{c^J}\eta
   ~~~~~~~\eta^+_I =  \int\limits_{c_I}\eta^+
   ~~~~~~~\eta^{+J} =  \int\limits_{c^J}\eta^+
   ~.$$
and finally from the two forms fields
       $$ x^+_\mu =\int\limits_{\Sigmag} X^+_\mu~,~~~~~~~~~~b^{+\mu} =\int\limits_{\Sigmag}\beta^{+\mu}~.$$

The global structure, {\it i.e.} the change of coordinates under the change
 $y^i=y^i(x)$ can be obtained from (\ref{change_coordinates_components}). One explicitly gets
\begin{equation}\label{change_coordinates_finite}
\eta^{+i}_I = \frac{\partial y^i}{\partial x^\mu} \eta^{+\mu}_I ~,~~\eta^{+iI} =
 \frac{\partial y^i}{\partial x^\mu} \eta^{+\mu I} ~,~~~ b_i =
 \frac{\partial x^\mu}{\partial y^i} b_\mu ~,~~~b^{+i}=
 \frac{\partial y^i}{\partial x^\mu} b^{+\mu}+
 \frac{\partial^2 y^i}{\partial x^\mu \partial x^\nu} \eta^{+\mu}_I\eta^{+\nu I}~,
\end{equation}
$$
\eta_{Ii} = \frac{\partial x^\mu}{\partial y^i} \eta_{I\mu} +
\frac{\partial}{\partial x^\nu}\frac{\partial x^\mu}{\partial y^i}\eta^{+\nu}_I b_\mu ~,~~~
\eta_{i}^I = \frac{\partial x^\mu}{\partial y^i} \eta_{\mu}^I + \frac{\partial}{\partial x^\nu}
\frac{\partial x^\mu}{\partial y^i}\eta^{+I\nu} b_\mu~,
$$
$$
x^+_i = \frac{\partial x^\mu}{\partial y^i} x^+_\mu -
 \frac{\partial}{\partial x^\nu}\frac{\partial x^\mu}{\partial y^i}(\eta^{+\nu}_I\eta^I_\mu -
  \eta^{+I\nu}\eta_{I\mu}) - \frac{\partial}{\partial x^\nu}\frac{\partial x^\mu}{\partial y^i}
  b^{+\nu} b_\mu - \frac{\partial^2\ }{\partial x^\nu\partial x^\rho}\frac{\partial x^\mu}{\partial y^i}
  \eta^{+\nu}_I\eta^{+\rho I} b_\mu \;.
$$

All the BV structure goes to the quotient which is a finite dimensional $BV$ manifold.
The odd symplectic structure (\ref{oddsymplecticsty}) reads
       \beq
      \omega = d x^\mu d x^+_\mu +   d\eta_{I\mu} d\eta^{+I\mu}  -     d\eta^I_{\mu} d\eta^{+\mu}_I
       + db_\mu db^{+\mu}~.
     \eeq{oddbecome002888}
Moreover, the BV action $S_{BV}$ defined in (\ref{BVfullaction}) when restricted to
 the constraint surface depends only on the reduced variables, {\it i.e.} it is a pull-back of a function
 on the reduced manifold. We use the same notation $S_{BV}$ for the reduced action.
 The reduced BV  action is computed from (\ref{BVfullaction}) as
$$ S_{BV} =   \alpha^{\mu\nu} (x)
 \eta_{I\mu} \eta^I_\nu  +  x^+_\mu \alpha^{\mu\nu}(x) b_\nu
 - \eta^{+\mu}_I  \d_\mu \alpha^{\rho\nu}(x) \eta^I_\rho b_\nu+$$
\beq
 +  \eta^{+I\mu} \d_\mu \alpha^{\rho\nu}(x) \eta_{I\rho} b_\nu
  -\frac{1}{2} b^{+\mu} \partial_\mu \alpha^{\rho\nu}(x)b_\rho b_\nu
   - \frac{1}{2} \eta^{+\mu}_I  \eta^{+I\nu} \partial_\mu \partial_\nu \alpha^{\rho\sigma} (x)b_\rho b_\sigma ~ ,
  \eeq{BVactionfindikksk}
 which obviously satisfies the classical master equation.

\medskip
The reduced variables can be assembled in the superfields ${\underline\Phi}=
(\superx^\mu,\supereta_\mu)$
\begin{equation}
\label{reduced_superfields}
 \superx^\mu = x^\mu + e^J \eta^{+\mu}_J + e_J \eta^{+J\mu} - s b^{+\mu} \;,
  ~~~~ \supereta_\nu = b_\nu + e_I \eta^I_\nu + e^I \eta_{I\nu} + s x^+_\nu\;,
\end{equation}
where $s$ is the generator of $H^2_{dR}(\Sigmag)$ normalized to $\int\limits_\Sigma s =1$.
The ring structure for $g>0$ is  defined by $e^J\wedge e_I = \delta_I^J s$ and by
 $s^2=0$ for $g=0$. One can check that the transformations of coordinates
  (\ref{change_coordinates_components}) can be deduced from the transformations of superfields
$$
\supery^i =y^i(\superx) ~,~~~~~ \supereta_i = \frac{\partial x^\mu}{\partial y^i}(\superx) \supereta_\mu\;.
$$

The action (\ref{BVactionfindikksk}) can be written as
\begin{equation}
\label{fin-dim-AKSZaction}
S_{BV} = \int ds ~ \alpha^{\mu\nu}(\superx)\supereta_\mu \supereta_\nu \;,
\end{equation}
where $\int ds$ is the induced integral on $H^\bullet_{dR}(\Sigmag)$.
 We easily see that this is an AKSZ formulation of the reduced theory,
 where we take as ``coordinates'' on the source the generators of the ring
  $H_{dR}(\Sigmag)$. Compared with the discussion in Section \ref{AKSZgeneral},
  here coordinates $\{e_J,e^J\}$ on the source for $g\not=1$  are not free but
   simply generate the commutative graded algebra $H_{dR}(\Sigmag)$.
   We may regard it as the sheaf $X_\Sigmag$ obtained by putting over a point
   the commutative graded algebra $H_{dR}(\Sigmag)$ such that that the reduced
   manifold is the space ${\rm Maps}(X_\Sigmag,T^*[1]M)$ of morphisms
    between $X_\Sigmag$ and $T^*[1]M$. The case $g=1$ is special since
     the cohomology ring is freely generated and we have a true graded
     manifold $X_{\Sigma_1}=\R^2[1]$. We will comment more on this in Section \ref{summary}.

\begin{remark}\label{grdmfld2d}{\rm
For $g=0$, ${\rm Maps}(X_{\S^2}, T^*[1]M)$ can be equivalently described as
 $T^*[-1]T^*[1]M$, see \cite{Bonechi:2007ar}.  For $g>0$ the BV manifold can be described as
 $${\rm Maps}(X_{\Sigmag}, T^*[1]M) = T^*[-1] \left ({\rm Maps} (X_{L}, T^*[1]M)\right )~,$$
  where $X_{L}$ is the subsheaf over a point generated by $(1, e_I)$. The total dimension of
   the BV manifold is given by the following formula
  $$ \dim~ {\rm Maps}(X_{\Sigmag}, T^*[1]M) =  4(g +1)  \dim M~.  $$
  }
  \end{remark}

Since the reduced BV-manifold is finite dimensional we can introduce an integral.
Let $\Omega=\rho_\Omega ~dx^1\ldots dx^n$ be a volume form on $M$. It defines
the generator of the Schouten bracket $D_\Omega$ on $\Lambda TM = C^\infty(T^*[1]M)$.
Moreover it defines a Berezinian integration on the reduced BV manifold
 ${\rm Maps}(X_\Sigmag,T^*[1]M)$ as
$$\mu_\Omega = \rho_\Omega^{2(2-2g)} Dz~,$$
where $Dz=dx\ldots dx^+\ldots db\ldots\ldots db^+\ldots d\eta \ldots d\eta^+$ is the coordinate
 volume form.
Since under the change of coordinates $\tilde{z}=\tilde{z}(z)$ the coordinate volume form transforms as
\begin{equation}\label{berezinian}
D\tilde{z}={\rm Ber}\frac{\partial \tilde z}{\partial z} Dz~, ~~~~ {\rm Ber}\left(\begin{array}{cc}I_{00}& I_{01}\cr I_{10}&I_{11}\end{array}\right)=\frac{\det(I_{00}-I_{01}I_{11}^{-1}I_{10})}{\det I_{11}}\end{equation}
it is a tedious but straightforward computation to put (\ref{change_coordinates_finite}) in
(\ref{berezinian}) and check that $\mu_\Omega$ is well defined.

The corresponding generator of the BV-bracket is
\begin{equation}\label{generatorBVgenus}\Delta_\Omega= \frac{\d}{\d x^+_\mu}\frac{\d}{\d x^\mu}  -
 \frac{\d}{\d b^{+\mu}}\frac{\d}{\d b_\mu}
-  \frac{\d}{\d \eta^{+\mu}_I}\frac{\d}{\d \eta^I_\mu} +
 \frac{\d}{\d \eta^{+I\mu}}\frac{\d}{\d \eta_{I\mu}}
 + (2-2g)\{\log\rho_\Omega,-\}\;. \end{equation}
We then compute
 $$ \Delta_\Omega S_{BV} = (g-1) (D_\Omega \alpha)^\mu b_\mu=(g-1)\chi_\Omega^\mu b_\mu~,$$
where $\chi_\Omega=D_\Omega \alpha$ is the modular vector field that satisfies
 $d_{LP}\chi_\Omega=0$. The class $[\chi_\Omega]\in H_{LP}(M;\alpha)$ is called {\it modular class} and
  it is independent from the choice of the volume form $\Omega$.  The modular class represents an obstruction for
  $S_{BV}$ to satisfy the quantum master equation for any genus $g\not=1$. This extends to
  any genus the construction in \cite{Bonechi:2007ar}; the result has been already observed
  in \cite{Cattaneo:2008yf}. In order to understand from the point of view of QFT this dependence
   on the genus, we recall the renormalization of the PSM on the disk in \cite{Cattaneo:1999fm};
    in fact in the expansion on Feynman graphs, the modular vector field appeared as a factor of the
    divergent graphs with self insertions. Their renormalization was obtained by introducing a non
    vanishing vector field, which in the compact case is possible only for $g=1$.

Any $w=w^{\mu_1\ldots \mu_k}\beta_{\mu_1}\ldots \beta_{\mu_k}$ representing a class in
 Lichnerowicz-Poisson cohomology defines the chain of observables
 $\Phi^* w=w^{\mu_1\ldots \mu_k}({\mathsf X}){\boldsymbol{\eta}_{\mu_1}}
  \ldots{\boldsymbol{\eta}_{\mu_k}}=O^{(0)}(w)+ O^{(1)}_{\alpha}(w)\theta^\alpha+\ldots$
   of the full theory. They are invariant with respect to gauge transformations (\ref{gauge_transf_superfield})
   and correspond to the observables ${\underline\Phi}^*w=w^{\mu_1\ldots \mu_k}(\superx)\supereta_{\mu_1}\ldots \supereta_{\mu_k}=O^{(0)}(w)+ O^{(1)}_{I}(w)e^I+\ldots$ of the reduced AKSZ theory.
   By using (\ref{generatorBVgenus}) we compute the following relation
$$
\Delta_\Omega {\underline\Phi}^*w= 2s (1-g) {\underline \Phi^*} D_\Omega w\;.
$$

We conclude that, given a unimodular Poisson structure $\alpha$, for any genus $g\not=1$
the finite dimensional BV theory maps any class $[w]\in H_{LP}(M;\alpha)$ represented by
 a divergenceless vector field $w$ to the class represented by $e^{S_{BV}}{\underline\Phi}^*w$ in the cohomology of the complex ($C^\infty({\rm Maps}(X_\Sigmag,T^*[1]M)),\Delta_\Omega$).  For $g=1$ the hypothesis of unimodularity is not needed and such invariants are defined for any class in Lichnerowicz Poisson cohomology. The main point is now to characterize these invariants in terms of more canonical cohomologies and mainly understand when they are not trivial. Here we simply sketch some obvious considerations and leave a more systematic analysis for further investigations.

One has to compute the correlators of such observables by choosing a
gauge fixing, {\it i.e.} a lagrangian submanifold $\cal L$. There is always
 one canonical choice which consists simply in putting all antifields equal
 to zero, {\it i.e.} $x^+=\eta^+=b^+=0$. The case $g=0$ has been studied
  in \cite{Bonechi:2007ar} and we refer to it for the result. The case $g>0$
   is badly defined due to fibrewise integration along $\eta$ in the degenerate
   directions of the Poisson tensor.

If one allows on the target $M$ a complex structure and introduces complex
coordinates $\{z^a,z^\abar\}$, then one can define a gauge fixing by putting
$x^+=b^+=0$ and $\eta_{Ia}=\eta^I_{\abar}=\eta^{+I\abar}=\eta^{+a}_I=0$.
By looking at (\ref{change_coordinates_components}) one can easily check
that this is invariant under holomorphic transformation of variables. In the
 symplectic case the computation of the partition function in this gauge fixing
  gives the Euler number of $M$; due to degeneracy of the Poisson tensor it
  is not clear how to give a meaning to this integral in the general Poisson case.

\section{Three dimensional case: Courant sigma model}
\label{CSM}

In this Section we consider the reduction of three dimensional topological theory associated to any
 Courant algebroid,  the Courant sigma model. We follow closely the presentation given in  \cite{Roytenberg:2006qz}
  (see also for the related discussion \cite{Ikeda:2002wh, Hofman:2002rv, Hofman:2002jz}).

We follow notations of Section 4 of \cite{Roytenberg:2006qz}. Let $E\rightarrow M$
 be a vector bundle equipped with a fiberwise nondegenerate symmetric inner
  product $\langle,\rangle$, of arbitrary signature.
A {\it Courant algebroid} structure on $(E,\langle,\rangle)$ is a a bilinear operation
 $\circ$ on sections of $E$ and a bundle map (the {\it anchor}) $a:E\rightarrow TM$
  satisfying the following properties
\begin{itemize}
 \item [$i$)] $s\circ (s_1\circ s_2)=(s\circ s_1)\circ s_2 + s_1\circ(s\circ s_2)$;
 \item[$ii$)] $a(s_1\circ s_2)=[a(s_1),a(s_2)]$;
 \item[$iii$)] $s_1\circ(f s_2)= f(s_1\circ s_2)+(a(s_1)(f)) s_2$;
 \item[$iv$)] $\langle s, s_1\circ s_2+s_2\circ s_1\rangle = a(s)(\langle s_1,s_2\rangle)$;
 \item[$v)$] $a(e)(\langle s_1,s_2\rangle) = \langle s\circ s_1,s_2\rangle + \langle s_1, s\circ s_2\rangle$,
\end{itemize}
for $s,s_1,s_2\in \Gamma(E)$ and $f\in C^\infty(M)$.

We can associate to these data a symplectic graded manifold $(\M,\omega)$ with
 $\deg\omega=2$. Locally we introduce a set of coordinates $\{x^\mu\}$
 on $M$ and the local basis $\{e_a\}$ of sections of $E$ such that $\langle e_a,e_b\rangle = g_{ab}$,
  with constant $g_{ab}$. Then the symplectic non negatively graded manifold $(\M,\omega)$
   is described in terms of  the local coordinates  $\{x^\mu,p_\mu,\xi^a\}$, with $\deg(x)=0$,
   $\deg(p)=2$ and $\deg(\xi)=1$ with the symplectic form of degree $2$
$$
\omega = dp_\mu dx^\mu + \frac{1}{2} d\xi^a g_{ab} d\xi^b~.
$$
Globally $\M$ can be interpreted as the symplectic submanifold of $T^*[2] E[1]$ defined by $\theta_a = \frac{1}{2} g_{ab} \xi^b$
 with $\theta$ being a momentum for $\xi$.
 As shown in \cite{Roytenberg:2002nu}  the
 Hamiltonian function of degree 3
\begin{equation}\label{hamiltonian_courant}
\Theta = \xi^a P_a^\mu p_\mu - \frac{1}{6} T_{abc} \xi^a\xi^b\xi^c~,
\end{equation}
where  the structure constants $T_{abc}=\langle e_a\circ e_b, e_c\rangle$ and the anchor
 $a(e_a)=P_a^\mu \partial_\mu$ defines on $(E,\langle,\rangle)$
 a {\it Courant algebroid} structure if and only if $\{\Theta,\Theta\}=0$.
 Due to the AKSZ logic reviewed in Section \ref{AKSZgeneral} one can define a three dimensional theory
whose space of fields is ${\rm Maps}(T[1] \Sigma, \M)$ with $\Sigma$ being any three dimensional manifold.
     Introducing the superfields
  $({\mathsf X}^\mu,   \boldsymbol{P}_\mu,  \boldsymbol{\xi}^a)$ corresponding to the local description
   of $\M$  the Courant sigma model is defined by the following BV action
 \beq
  S_{BV} = \int d^3\theta d^3 u \left (  \boldsymbol{P}_\mu D {\mathsf X}^\mu + \frac{1}{2} \boldsymbol{\xi}^a g_{ab}
   D \boldsymbol{\xi}^b
   - \boldsymbol{\xi}^a P_a^\mu \boldsymbol{P}_\mu + \frac{1}{6} T_{abc} \boldsymbol{\xi}^a
   \boldsymbol{\xi}^b \boldsymbol{\xi}^c
   \right )~.
 \eeq{Courantaction}
The AKSZ construction guarantees that (\ref{Courantaction}) is a solution of the classical master equation associated to
  any three dimensional manifold $\Sigma$ and the Courant algebroid $E$.
 The reader may consult  \cite{Roytenberg:2006qz} for the explicit expression of this $BV$ action
 in components of the superfield.

   Here we are interested in the finite dimensional BV theory obtained through the  symplectic reduction of
    the AKSZ theory described above. The whole construction is done in complete analogy with the two dimensional
    case,  see subsection \ref{reduction_cohomology}. Thus we skip the details of the actual reduction and
     present the final answer.  Introduce the basis in $H_{dR}(\Sigma)$ such that $\{e_I\}$ is a basis in $H^1_{dR} (\Sigma)$,
   $e^I$ is the  dual basis in $H^2_{dR} (\Sigma)$, $1$ basis element in $H^0_{dR}(\Sigma)$ and $s$ is a basis element in
       $H^3_{dR}(\Sigma)$.  With this choice the natural  ring structure on $H_{dR}(\Sigma)$ is given
 $$ e_I \wedge e^J = \delta^I_J s~,~~~~e_I \wedge e_J = f_{IJK} e^K$$
 and also  it follows that
   $$ e_I \wedge e_J \wedge e_K = f_{IJK} s~.$$
  The constants  $f_{IJK}$ have the interpretation as intersection numbers of one cycles.
   Define the sheaf $X_\Sigma$ by putting over a point the commutative graded algebra $H_{dR}(\Sigma)$. The space of maps
    ({\it i.e.} morphisms of ringed spaces)  ${\rm Maps}(X_{\Sigma}, \M)$ is equipped with an odd symplectic structure. We can introduce
    the superfields expanded in $(1, e_I, e^I, s)$
\ber
\nonumber {\mathsf X}^\mu& =& x^\mu + F^{+\mu I} e_I  + \alpha^{+\mu}_I e^I + \gamma^{+\mu} s~,\\
\label{reducedsuperfields3d}\boldsymbol{\xi}^a &= &\beta^a + A^{aI} e_I + g^{ab} A^{+}_{bI} e^I + g^{ab} \beta^{+}_b s~,\\
\nonumber  \boldsymbol{P}_\mu &=& \gamma_\mu + \alpha_\mu^I e_I + F_{\mu I} e^I + X^+_\mu s~,
 \eer{12345}
 which correspond to the local description of element from ${\rm Maps}(X_{\Sigma}, \M)$ .
 The integration is naturally defined by the relation $\int ds~ s = 1$ and all other integrals are zero (modulo ring relations).
  The odd symplectic structure is
  \beq
  \omega = \int ds~ \left ( \delta {\mathsf X}^\mu \delta \boldsymbol{P}_\mu + \frac{1}{2} \delta \boldsymbol{\xi}^a
  g_{ab} \delta \boldsymbol{\xi}^b\right )~.
  \eeq{sympdp39938}
Upon the reduction the action (\ref{Courantaction}) gives rise to the following BV action
\ber
\nonumber & &S_{BV} = - A^{aI} P_a^\nu(x) F_{\mu I} + \frac{1}{6} T_{abc}(x) f_{IJK} A^{aI} A^{bJ} A^{cK} - \beta^a P_a^\mu (x) X^+_\mu \\
\nonumber & &+  (- g^{ac} P_a^\mu(x) \alpha^I_\mu -   \beta^a A^{bI} T_{abr} g^{rc} )A_{cI}^+\\
\nonumber & & +  (-\beta ^{a}\partial _{\mu}P^{\nu}_{a}F_{\nu I} +f_{JKI}A^{a J}\partial _{\mu}P^{\nu}_{a}\alpha^K _{\nu}+\frac{1}{2}f_{JKI}\partial _{\mu}T_{abc}\beta ^{a}A^{bJ}A^{cK})F^{+\mu I} \\
 \label{reducedBV3d}  & &+(-\beta ^{a}\partial _{\mu}P^{\nu}_{a}\alpha^I _{\nu}-A^{aI}\partial _{\mu}P^{\nu}_{a}\gamma _{\nu}+\frac{1}{2}\partial _{\mu}T_{abc}\beta ^{a}\beta ^{b}A^{cI})\alpha ^{+\mu}_I\\
 \nonumber & & +(\frac{1}{2}T_{abr}g^{rc}\beta ^{a}\beta ^{b}-g^{ac}P_{a}^{\mu}\gamma _{\mu})\beta ^{+}_{c}+(\frac{1}{6}\partial _{\mu}T_{abc}\beta ^{a}\beta ^{b}
\beta ^{c}-\beta ^{a}\partial _{\mu}P^{\nu}_{a}\gamma _{\nu})\gamma ^{+\mu}\\
\nonumber & &- (\frac{1}{2}\partial _{\mu}T_{abr}g^{rc}\beta ^{a}\beta ^{b}-g^{ac}\partial _{\mu}P^{\nu}_{a}\gamma _{\nu})F^{+\mu I}A^{+}_{cI}+\frac{1}{2}(\beta^{a}
 \partial _{\mu}\partial _{\nu}P^{\rho}_{a}\alpha^K _{\rho}+A^{aK }\partial _{\mu}\partial _{\nu}P^{\rho}_{a}\gamma _{\rho}\\
\nonumber    & & -\frac{1}{2}\partial _{\mu}\partial _{\nu}T_{abc}\beta ^{a}\beta ^{b}A^{cK})F^{+\mu I}F^{+\nu J}  f_{KIJ}
     -(\beta ^{a}\partial _{\mu}\partial _{\nu}P^{\rho}_{a}
 \gamma _{\rho}-\frac{1}{6}\partial _{\mu}\partial _{\nu}T_{abc}\beta ^{a}\beta ^{b}\beta ^{c})F^{+\mu I}  \alpha^{+\nu}_I\\
\nonumber &&- \frac{1}{6}(-\beta ^{a}\partial _{\mu}\partial _{\nu}\partial _{\rho}P^{\sigma}_{a}\gamma _{\sigma}+\frac{1}{6}\partial _{\mu}\partial _{\nu}\partial _{\rho}T_{abc}\beta ^{a}\beta ^{b}
 \beta ^{c})F^{+\mu I} F^{+\nu J} F^{+\rho K} f_{IJK}~,
\eer
 which automatically satisfies the classical master equation if $E$ is equipped with the structure of a Courant algebroid.

 On the given BV-manifold ${\rm Maps}(X_\Sigma,\M)$ the Berezinian integration can be defined as
\begin{equation}
\label{berezinian3d}
\mu = g^{1-b_1} Dz ~,
\end{equation}
where $b_1=\dim H^1_{dR}(\Sigma)$, $Dz$ is  the canonical volume form with respect to the coordinates
 introduced in (\ref{reducedsuperfields3d}) and  $g=\det(g^{ab})$.
 The Berezinian integration is canonically defined (see Appendix \ref{ap-ber} for the details).
 Thus the corresponding generator (odd Laplacian) of the BV
   bracket coincides with the naive one
\beq
\Delta_\mu= \frac{\d}{\d x^+_\mu}\frac{\d}{\d x^\mu} - \frac{\d}{\d \beta^+_a}\frac{\d}{\d \beta^a} +\frac{\d}{\d A^{a I}}\frac{\d}{\d A^{+}_{aI}} + \frac{\d}{\d \gamma^{+\mu}}\frac{\d}{\d \gamma_\mu} - \frac{\d}{\d \alpha^{+\mu}_I}\frac{\d}{\d \alpha^I_\mu} - \frac{\d}{\d F^{+\mu I}}\frac{\d}{\d F_{\mu I}}\;
\eeq{Laplaciancour}
Using the explicit expression (\ref{reducedBV3d}) and the axioms of Courant algebroid we easily check that
$$
\Delta_\mu S_{BV} = 0\;,
$$
{\it i.e.} $S_{BV}$ satisfies the quantum master equation.

As an example we can consider the case when the source manifold $\Sigma$ is
 a three sphere $\S^3$. In this case the reduced BV manifold ${\rm Maps}(X_\Sigma,\M)$
  corresponds to $T^*[-1]\M$ since $H_{dR}(\S^3)$ has only elements of degree $0$ and $3$.
   To describe the reduced BV theory for $\S^3$  we have to set  fields $A^{aI}$, $\alpha_\mu^a$, $F_{\mu I}$
    together with  their antifields $A_{bI}^+$, $\alpha_I^{+\mu}$, $F^{+\mu I}$ to zero in (\ref{reducedsuperfields3d}), (\ref{sympdp39938}),
     (\ref{reducedBV3d}) and remove them from the measure (\ref{berezinian3d}) and Laplacian (\ref{Laplaciancour}).
    It is easy to describe the observables and the correlators in this theory.
      The Hamiltonian (\ref{hamiltonian_courant}) defines
 an homological vector field $Q=\{\Theta,-\}$. The complex $(C^\infty(\M)_{pol}, Q)$
  of functions polynomial in $p$ is called the {\it standard complex} and we denote
   its cohomology with $H_Q(\M)_{pol}$.  However in what follows we need to consider
    the complex $(C^\infty(\M)_{exp}, Q)$ of functions with exponential decay in $p$ directions.
     We have to use  this complex in order to make sense of the integrals.
   For $f\in C^\infty(M)_{exp}$, $Q(f)=0$ the corresponding observable is
$$
\Phi^*(f) = O^{(0)}(f) + s O^{(3)}(f)~,
$$
which satisfies by construction
$$
\delta_{BV}(\Phi^*(f)) = \{S_{BV},\Phi^*(f)\}=0~,
$$
  where $\Phi \in {\rm Maps}(X_{\S^3},\M)$.  For $T^*[-1]\M$ we
   choose $\M$ as a Lagrangian submanifold defined by setting
all  antifields to zero, {\it i.e.} $X^+=\gamma^+=\beta^+=0$. Thus on $\M$ we have that $\Phi^*(f)= O^{(0)}(f)=f$
 and  the berezinian measure on $\M$ reads
$$
\sqrt{\mu} = \sqrt{g}\ dx\ d\gamma\ d\beta~.
$$
The correlator is defined as integral over the Lagrangian submanifold
\begin{equation}\label{correlation_function_courant}
\langle \Phi^*(f)\rangle = \int\limits_{\M} \sqrt{\mu}\ f\;.
\end{equation}
 To make this integral well-defined we have to assume that $f$ decays fast enough along non-compact
  directions. By construction the correlator (\ref{correlation_function_courant}) satisfies the version of
   the Stokes theorem with respect to BV-differential and odd Laplacian.

\medskip
\begin{example} {\rm Consider  $E$ to be a  Lie algebra with invariant metric interpreted  as a vector
 bundle over a point. In this case the Courant sigma model is just standard Chern-Simons theory.
  The reduced finite dimensional theory is BV version of the matrix theory.
  The standard complex is just $\Lambda E^*$ equipped with the Chevalley-Eilenberg differential. The  integral defined in (\ref{correlation_function_courant})
is different from zero only on top forms.}
 \end{example}

\begin{example}
 {\rm Consider $E=TM+T^*M$ with the canonical inner product and the Dorfman bracket
$$
[X_1\oplus \omega_1,X_2\oplus \omega_2]= [X_1,X_2]  \oplus L_{X_1}\omega_2
- \iota_{X_2} d\omega_1 \;.$$
The coordinates on the fiber now split  $\xi \rightarrow \{\psi^\mu,\theta_\mu\}$ and
 the corresponding graded manifold is described as the even symplectic manifold
 $\M=T^*[2]T[1]M$. The Hamiltonian (\ref{hamiltonian_courant}) reads $\Theta = \psi^\mu p_\mu$ giving
  rise to the
  homological vector field
$$Q = \psi^\mu \frac{\partial}{\partial x^\mu} + p_\mu \frac{\partial}{\partial \theta_\mu}~. $$
As shown in \cite{Roytenberg:2002nu}, $Q$ commutes with
$$\epsilon=p_\mu\frac{\partial}{\partial p_\mu}+ \theta_\mu\frac{\partial}{\partial\theta_\mu}\;.$$
The complex of polynomial functions in $p$ $(C^\infty(\M)_{pol},Q)$ then decomposes
 according to the $\epsilon$-degree, {\it i.e.} $C^\infty(\M)_{pol}=$  $\oplus_{k\geq 0} C^\infty(\M)_{(k)}$,
 with the subcomplex of degree zero is $(C^\infty(\M)_{(0)}, Q)=(\Omega (M),d)$ being
 the de Rham complex for $M$. Moreover, since $\epsilon= Q\iota+\iota Q$ with
  $\iota=\theta_\mu\frac{\partial}{\partial p_\mu}$  the cohomology $H_Q(\M)_{pol}$
   is concentrated in degree zero and isomorphic to the de Rham cohomology $H_{dR}(M)$.
    However for the BV theory we need the complex $(C^\infty(\M)_{exp}, Q)$ of functions with
     the exponential decay in $p$.  It is not clear how the corresponding cohomology is related to
      $H_{dR}(M)$. Otherwise the correlators can be defined in the way we described above. }
\end{example}

\section{Two dimensional case with boundary}
\label{boundary}

In this section we discuss the reduction of PSM on the surface with a boundary.
 Take a surface $\Sigmagn$, of genus $g$ and $n$ boundary components. We  consider
  the boundary conditions for PSM that have been introduced in \cite{Cattaneo:2003dp}.
  Let $\partial_i\Sigmagn$ the $i$-th boundary component of $\Sigmagn$ and let $C_i$
  a coisotropic submanifold of the Poisson manifold $(M,\alpha)$. We recall that a submanifold
  $C$ of $M$ is coisotropic if $\alpha(N^*C)\subset TC$, where $N^*C$ is the conormal bundle
  of $C$. We assume that the superfields $(\superX,\superEta)$ restrict to
  ${\rm Maps}(T[1]\partial_i\Sigma,N^*[1]C_i)$ for any $i=1,\ldots n$. More general boundary conditions
   have been introduced in \cite{Calvo:2005re}, where it is required that the rank of
    $\alpha(N_x^*C)+T_xC\subset T_xM$ is constant as $x$ varies over $C$.

We apply the same construction as in the closed case, {\it i.e.} we perform the reduction with
respect to the constraints defined in (\ref{constraints_finite}). Since the bracket between the constraints has now
 a boundary contribution
$$
\{\Lambda , T \} = \int\limits_{T[1]\partial\Sigma} \Lambda_\mu DT^\mu~,
$$
 we require that
\begin{equation}
\Lambda|_{\partial_i\Sigmagn} \in \Gamma(\Lambda T^*\partial_i\Sigmagn\otimes X^*N^*C_i)
\;,~~~~~ T|_{\partial_i\Sigmagn} \in\Gamma(\Lambda T^*\partial_i\Sigmagn\otimes X^*TC_i)~.
\label{boundary_constraints}
\end{equation}
in order to have a consistent reduction.
Due to the presence of the boundary conditions for fields and constraints, cohomologies
of $\Sigmagn$ relative to boundary components will appear in the description of the reduced BV-manifolds.
 In Appendix \ref{relative_homology}  we collect the relevant facts about relative (co)homology.

\subsection{The case with one boundary component}

 Consider the case of a surface $\Sigmaguno$ with one boundary component with the boundary condition
  corresponding to a coisotropic submanifold $C$. Let us introduce a set of coordinates of $M$ adapted to $C$ with $\{x^a\}$
tangent to $C$ and $\{x^n\}$ normal. Coisotropy of $C$ is then simply expressed by the condition $\alpha^{mn}=0$.

The gauge transformations (\ref{gauge_transf_superfield}) defined by the constraints together
with boundary conditions (\ref{boundary_constraints}) imply that the reduced BV manifold is
described by the following variables
$$
\superX^a\in H_{dR}(\Sigma_{g,1}) \;, ~~~~ \superX^n \in H_{dR}(\Sigma_{g,1},\partial\Sigma_{g,1})\;,~~~~
\superEta_a \in H_{dR}(\Sigma_{g,1},\partial\Sigma_{g,1}) \;,~~~ \superEta_n \in H_{dR}(\Sigma_{g,1})\;.
$$
The covariant meaning of the above statements and the gluing data of the reduced BV-manifold are better understood once that we introduce the reduced variables.

We recall from Appendix \ref{relative_homology} that $H^2(\Sigma_{g,1}) = H^0(\Sigma_{g,1},\partial\Sigma_{g,1})=0$.
In order to define the reduced coordinates, let us choose a set of representatives defining a basis of the relevant homologies. Let $u_0\in\Sigma_{g,1}$ be a representative generating $H_0(\Sigma_{g,1})$, $\{c^I\}$ for $H_1(\Sigma_{g,1})$, $\{g^{I'}\}$ for $H_1(\Sigma_{g,1},\partial\Sigma_{g,1})$ and the whole surface $\Sigma_{g,1}$ for $H_2(\Sigma_{g,1},\partial\Sigma_{g,1})$. Let $\{c_I\}$ and $\{g_{I'}\}$ be the dual basis for $H^1_{dR}(\Sigma_{g,1})$ and $H^1_{dR}(\Sigma_{g,1},\partial\Sigma_{g,1})$ respectively. Remark that beyond the natural pairings between homology and cohomology one can pair $H_1(\Sigma_{g,1})$ with $H^1_{dR}(\Sigma_{g,1},\partial\Sigma_{g,1})$. The matrix
$$\lambda^{I}_{I'}=\int\limits_{c^I}g_{I'}$$
 describes the natural homomorphism $H^1_{dR}(\Sigma_{g,1},\partial\Sigma_{g,1})\rightarrow H^1_{dR}(\Sigma_{g,1})$ as $g_{I'}\rightarrow \lambda^{I}_{I'} c_I$. Finally we define the following matrices
$$B^{(g,1)}_{I'J'}=\int\limits_{\Sigma_{g,1}} g_{I'}\wedge g_{J'}~,~~~~~~~~~A^{(g,1)}_{II'}=\int\limits_{\Sigma_{g,1}} c_I\wedge g_{I'}~,$$
  where $A^{(g,1)}$ is non degenerate matrix.

Then we define
$$
x^a = X^a(u_0)~,~~~~ b_n = \beta_n(u_0)\;,
$$
$$
\eta_a^{I'} = \int\limits_{g^{I'}} \eta_a ~,~~~\eta_n^I =\int\limits_{c^I} \eta_n ~,~~~~
\eta^{+aI} = \int\limits_{c^I} \eta^{+a}  ~,~~~~ \eta^{+nI'} = \int\limits_{g^{I'}} \eta^{+n}
$$
$$
x^+_a = \int\limits_\Sigma X^+_a ~~,~~~    b^{+n} = \int\limits_\Sigma \beta^{+n}\;.
$$

The reduced odd symplectic form reads
\beq
      \omega = d x^a d x^+_a +   d\eta_a^{I'} d\eta^{+aI} A^{(g,1)}_{II'}  +  d\eta_{n}^I d\eta^{+nI'} A^{(g,1)}_{II'}
       + db_n db^{+n}~.
     \eeq{oddsymplformboundary}

If we take another adapted system of coordinates $\{y^\alpha=y^\alpha(x^a,x^n), y^\nu=y^\nu(x^a,x^n)\}$
 with $y^\alpha$ being the coordinates along $C$ and $y^\nu$ transverse to $C$, then the
   law for the corresponding transformation of the reduced coordinates can be derived from (\ref{change_coordinates_components}). We will not explicitly write it here; we simply remark that beyond the matrix $(\partial y/\partial x)$, it involves all matrices $A_{IJ'}^{(g,1)},B_{I'J'}^{(g,1)}$ and $\lambda_J^{J'}$ introduced above.

We give here a non canonical description of the reduced BV manifold that depends on the choice of a tubular neighborhood for $C$. We recall that a {\it tubular neighborhood} for $C$ is an embedding $j:NC\rightarrow M$, where $NC$ is the normal bundle of $C$ inside $M$, such that $j(C)=C$. Let $X_{\Sigma_{g,1}}$ be the sheaf obtained by putting over a point the commutative graded algebra $H_{dR}(\Sigma_{g,1})$.

\medskip
\begin{proposition}
For any choice of a tubular neighborhood of $C$ inside $M$, there exists an isomorphism between the reduced BV-manifold and
$T^*[-1]({\rm Maps}(X_{\Sigma_{g,1}},N^*[1]C))$. Such isomorphism is canonical in the case $g=0$, {\it i.e.} $\Sigma=D^1$, when the reduced $BV$ manifold is $T^*[-1] N^*[1] C$.
\end{proposition}
{\it Proof}.
The superfields for the graded manifold ${\rm Maps}(X_{\Sigma_{g,1}},N^*[1]C)$ are
$$\superx^a = x^a + c_I \eta^{+Ia}\;,~~~~~~ \supereta_n = b_n + c_I \eta^I_n\;. $$
One can easily check that the degree of the momenta are correct.
Let us choose the tubular neighborhood of $C$. A trivialization of $NC$ with $(t^\nu_n)$ as transition functions of $NC$, defines the atlas of adapted coodinates $\{x^a,x^n\}$, where $\{x^a\}$ are coordinates of $C$ and $\{x^n\}$ are coordinates on the fibre. The change of coordinates is given by $$y^\alpha=y^\alpha(x^a),~~~~y^\nu=t^\nu_n(x^a) x^n~~~~~.$$
It is tedious but straightforward to check that the rules of change of variables defined in ${\rm Maps}(X_{\Sigma_{g,1}},N^*[1]C)$ together with those of momenta
are the same that we get from (\ref{change_coordinates_components}). $\Box$

The  reduced $BV$ action reads as
\begin{eqnarray*}
S_{BV} &=& x^+_a\alpha^{an}b_n - \frac{1}{2} b^{+m}\partial_m\alpha^{np} b_n b_p + \frac{1}{2}\alpha^{ab}\eta^{I'}_a\eta^{J'}_b B_{I'J'}^{(g,1)} + \alpha^{am}\eta_a^{I'}\eta_m^J A_{I'J}^{(g,1)} +\cr
& & \partial_b\alpha^{na}\eta^{+bI}\eta_a^{J'} b_n A_{IJ'}^{(g,1)} + \partial_m\alpha^{na}\eta^{+mI'}\eta_a^{J'} b_n B_{I'J'}^{(g,1)} - \partial_m\alpha^{np}\eta^{+mI'}\eta_p^{J} b_n A_{JI'}^{(g,1)} +\cr
& & -\frac{1}{4}\partial_p\partial_q\alpha^{mn}b_m b_n\eta^{+pI'}\eta^{+qJ'} B^{(g,1)}_{I'J'}-
\frac{1}{2}\partial_a\partial_p\alpha^{mn} b_m b_n \eta^{+aI}\eta^{pI'} A^{(g,1)}_{IJ'}\;~~~,
\end{eqnarray*}
 which automatically satisfies the classical master equation.

\begin{remark}{\rm  Apparently, there is no natural description of these solutions of BV equation as AKSZ-actions.}
\end{remark}

 Recall that $\dim H^1_{dR}(\Sigma_{g,n})=\dim H^1_{dR}(\Sigma_{g,n},\partial\Sigma_{g,n})=2g+n-1$.
  Any choice of volume form $\Omega=\rho_\Omega dx^a dx^n$ on $M$ defines the berezinian
$$
\mu_\Omega=\rho_\Omega^{2(1-2g)} dx^a dx^+_adb_n db^{+n}d\eta d\eta^+ $$
and the BV-generator
$$\Delta_\Omega= \frac{\d}{\d x^+_a}\frac{\d}{\d x^a}  -
 \frac{\d}{\d b^{+n}}\frac{\d}{\d b_n} + A^{(g,1)}{}^{I'J}\left(\frac{\d}{\d \eta^{I'}_a}\frac{\d}{\d \eta^{+aJ}}
-\frac{\d}{\d \eta^{J}_m}\frac{\d}{\d \eta^{+mI'}} \right) + (1-2g)\{\log\rho_\Omega,-\}\;. $$
We easily compute
$$
\Delta_\Omega S_{BV} = (1-2g)(\partial_a \alpha^{an} + \partial_m\alpha^{mn} + \partial_a\log\rho_\Omega \alpha^{an} ) b_n = (1-2g)\chi_{\Omega,N^*C}^n b_n~,
$$
where $\chi_{\Omega,N^*C} \in NC=T_CM/TC$ is a representative of the modular
  class of the Lie algebroid $N^*C$. We conclude that the solution of the classical
  master equation $S_{BV}$ satisfies the quantum master equation if and only if $N^*C$ is unimodular.

\section{Abstract AKSZ models from dg Frobenius algebras}\label{generalizationAKSZ}

Inspired by the previous consideration we suggest the extension of the AKSZ idea.  
The AKSZ construction admits the following straightforward generalization: loosely, one can use any differential graded (``dg'') Frobenius algebra (or a sheaf of dg  algebras such that the global sections carry the structure of the Frobenius algebra) instead of differential forms on the worldsheet. In this section we sketch two versions of the construction --- vector space version and sheaf version.

\subsection{Vector space version}\label{vector_space_version}

\textbf{\\The source:}
A unital dg Frobenius algebra $\C$, i.e. a $\ZZ$-graded vector space\footnote{Our convention is that \textit{elements} of $\C^i$ have degree $i$ (not coordinates on $\C^i$).} $\C=\C^0\oplus\cdots\oplus \C^{n+1}$ endowed with (super-)commutative associative multiplication $m:S^2\C\ra \C$ of degree 0, differential $D: \C^\bt\ra \C^{\bt+1}$ of degree 1 and non-degenerate symmetric pairing $\Pi: S^2\C \ra \RR$ of degree $-n-1$, satisfying the following axioms:
\begin{itemize}
\item Degree properties:
\begin{eqnarray*}
|m(u,v)|&=&|u|+|v| \\
|Du|&=&|u|+1 \\
\Pi(u,v)&\neq & 0 \;\mathrm{implies}\; |u|+|v|=n+1
\end{eqnarray*}
\item Symmetry properties:
\begin{eqnarray*}
m(u,v)&=&(-1)^{|u|\cdot |v|}m(v,u) \\
\Pi(u,v)&=&(-1)^{|u|\cdot |v|}\Pi(v,u)
\end{eqnarray*}
\item Poincar\'{e}, Leibniz and associativity identities for differential and multiplication:
\begin{eqnarray*}
D^2&=&0 \\
Dm(u,v)&=&m(Du,v)+(-1)^{|u|}m(u,Dv) \\
m(m(u,v),w)&=&m(u,m(v,w))
\end{eqnarray*}
\item Multiplication is cyclic w.r.t. the pairing:
$$\Pi(u,m(v,w))=\Pi(m(u,v),w)$$
\item Differential is skew-symmetric w.r.t. the pairing:
$$\Pi(Du,v)+(-1)^{|u|}\Pi(u,Dv)=0$$
\item Pairing $\Pi$ is non-degenerate, i.e. induces an isomorphism
$$\C^\bt\xrightarrow{\sim} (\C^{n+1-\bt})^*$$
\end{itemize}
Here we assume that $u,v,w\in \C$ are homogeneous elements and $|\cdots|$ denotes the degree of an element. Also, we denote the unit of $\C$ by $\1\in\C^0$.

Equivalently, one can describe $\C$ as a unital (super-)commutative dg algebra with trace $\Tr: \C^{n+1}\ra\RR$ (and extended by zero on lower degree components of $\C$) satisfying
$$\Tr(Du)=0$$
and such that the pairing $\Tr(m(\bt,\bt))$ is non-degenerate.
The trace is constructed from the pairing as $\Tr (u) =\Pi(\1,u)$ (and vice versa, pairing can be constructed from the trace as $\Pi(u,v)=\Tr(m(u,v))$\;).

\textbf{The target:} A $\ZZ$-graded vector space $\W$ endowed with a (constant) symplectic form $\omega\in S^2(\W[1])^*$ of degree $n$ and a function $\Theta\in S^\bt(\W^*)$ of degree $n+1$, satisfying $\{\Theta,\Theta\}=0$.

The space of BV fields of abstract AKSZ model is defined in this setting to be the $\ZZ$-graded vector space
\be\FF=\C\otimes \W~.\ee
Let $\{e_\zeta\}$ be a basis in $\C$ and $\{\tau_A\}$ be a basis in $\W$ (we denote the corresponding coordinates on $\W$ by $\{X^A\}$). Then $\{e_\zeta\otimes \tau_A\}$ is the basis in $\FF$ and we denote the corresponding coordinates on $\FF$ by $\{\Phi^{A\zeta}\}$. The degree (ghost number) of $\Phi^{A\zeta}$ is $-|e_\zeta|+|X^A|$. If the symplectic form $\omega$ on $\W$ is $\omega=dX^A \omega_{AB} dX^B$, then the degree -1 symplectic form $\Omega$ on $\FF$ is defined as
$$\Omega=\Tr(\delta\Phi^{A}\, \omega_{AB}\, \delta\Phi^{B})=(-1)^{(|A|+1)\cdot |\zeta|+n+1}\delta \Phi^{A\zeta}\; \Pi(e_\zeta,e_\eta)\;\omega_{AB}\;\delta \Phi^{B\eta}~,$$
where $\Phi^A:=\Phi^{A\zeta}e_\zeta$ (and we use multiplication $m$ under trace implicitly) and we use the obvious shorthand notation $|A|=|X^A|$, $|\zeta|=|e_\zeta|$.
The abstract AKSZ action is
\be S=S_{kin}+S_{int}=\underbrace{ \frac{1}{2}\Tr(\Phi^A\, \omega_{AB}\, D\Phi^B) }_{=(-1)^{|A|\cdot (|\zeta|+1)}\frac{1}{2}\Phi^{A\zeta}\;\Pi(e_\zeta,D e_\eta)\;\omega_{AB}\;\Phi^{B\eta}}+(-1)^{n+1}\Tr(\Phi^*(\Theta)) ~,\label{abstract AKSZ action}\ee
where $\Phi^*: S^\bt(\W^*)\ra \C$ is the ring homomorphism induced by the field $\Phi\in \C\otimes \W \cong \Hom(\W^*,\C)$ (i.e. we first interpret $\Phi$ as a map of graded vector spaces from $\W^*$ to $\C$ and then extend it as a ring homomorphism, according to free multiplication in $S^\bt(\W^*)$ and multiplication $m$ in $\C$). In coordinates: if
\be \Theta=\Theta_A X^A+\frac{1}{2} \Theta_{AB} X^A X^B+\cdots \label{Theta Taylor}\ee then
\be S_{int}=(-1)^{n+1}\Theta_A \Tr(e_\zeta)\;\Phi^{A\zeta}+(-1)^{(|A|+1)\cdot |\zeta|+n+1}\frac{1}{2}\Theta_{AB} \Tr(m(e_\zeta, e_\eta))\;\Phi^{A\zeta}\Phi^{B\eta}+\cdots \label{S_int Taylor}\ee

We claim that the action (\ref{abstract AKSZ action}) satisfies classical master equation $\{S,S\}=0$ w.r.t. the anti-bracket on $\FF$ associated to the odd symplectic form $\Omega$.

\begin{remark}{\rm The coordinate-free version of the construction above is as follows. We have a ring homomorphism $\ev^*: S^\bt(\W^*)\ra S^\bt(\FF^*)\otimes \C$ defined on generators as the canonical map $\W^*\ra \FF^*\otimes\C$ associated to the identity $\FF\ra \C\otimes\W=\FF$. (Notation $\ev^*$ should remind of the pull-back by evaluation map $\ev: \Maps(\N,\M)\times\N\ra \M$ in usual AKSZ construction which goes as $\ev^*: C^\infty(\M)\ra C^\infty(\Maps(\N,\M))\otimes C^\infty(\N)$.) The interaction part of action (\ref{abstract AKSZ action}) is then
$$S_{int}=(\id\otimes\Tr)\circ\ev^* (\Theta)~.$$
We can also formally extend $\ev^*$ to differential forms on $\W$ as a homomorphism of dg algebras
$\ev^*: \Omega^\bt(\W)\ra \Omega^\bt(\FF)\otimes \C$ (here $\C$ is treated as an algebra with zero de Rham differential). Then the odd symplectic form on $\FF$ is given by
$$\Omega=(\id\otimes \Tr)\circ\ev^*(\omega)~.$$
The kinetic part of action (\ref{abstract AKSZ action}) is defined as the Hamiltonian function for the cohomological vector field on $\FF$, induced by the differential $D:\C^\bt\ra \C^{\bt+1}$.}\end{remark}

\begin{example} (AKSZ with target a vector space.) {\rm Usual AKSZ models on the space $\Maps(\N,\M)$, in the case when $\M$ is a graded vector space with constant symplectic form, can be interpreted as abstract AKSZ with $\W=\M$ and $\C=C^\infty(\N)$.}\end{example}

\begin{example} (abstract Chern-Simons.) {\rm Taking arbitrary $\C$ with $n=2$ (i.e. concentrated in degrees 0,1,2,3 and with pairing of degree -3), taking $\W=\g[1]$ for a quadratic Lie algebra $\g$ with invariant pairing $\pi_\g$ and setting
$$\omega=dX^A\;\pi_\g(\tau_A,\tau_B)\; dX^B\quad,\quad \Theta=\frac{1}{6}\pi_\g(\tau_A,[\tau_B,\tau_C])\; X^A X^B X^C$$
we obtain abstract Chern-Simons in the sense of \cite{CM}.}\end{example}

In general, in this way (i.e. by allowing $\C$ to be an arbitrary dg Frobenius algebra with pairing of appropriate degree, instead of demanding that it is of form $C^\infty(\N)$) we can construct abstract versions of AKSZ models with vector space targets.

\subsection{Sheaf version}
We recall first some basic definition that can be found in \cite{Hartshorne}. A ringed space is a couple $(X,\O_X)$ where $X$ is a topological space and $\O_X$ is sheaf  of rings on $X$, see \cite{Hartshorne}. We denote with $\O_X(U)$ the local sections on the open $U\subset X$. A graded manifold $\M$ is a ringed space $(\M_0,\M)$ such that $\M(U)$ is locally isomorphic to $C^\infty(U)\otimes S(V^*)$, for some open $U\subset \M_0$ and graded vector space $V$.

\begin{definition}\label{morph_ring_space} A morphism of ringed spaces from  ($X,\O_X$) to ($Y,\O_Y$) is a couple ($\phi,\Phi$), where $\phi:X\rightarrow Y$ and  $\Phi: \O_Y\rightarrow \phi_* \O_X$ is a morphism of sheaves on Y.
\end{definition}

The pushforward sheaf is defined as the sheaf on $Y$ with local sections $\phi_* \O_X(U) = \O_X(\phi^{-1}U)$, for any open $U\subset Y$. The morphism ($\phi,\Phi$)  assigns to any open $U\subset Y$ a ring morphism  $\Phi^*(U): \O_Y(U)\rightarrow \O_X(\phi^{-1}U)$.

The abstract AKSZ construction depends on the following data.

\textbf{\\The source:} A sheaf $\J$ of dg supercommutative algebras over some closed manifold $\N_0$, such that locally, for some open $U\subset \N_0$, we have
\be \J(U)\cong \Omega^\bt(U)\otimes\C\;, \label{J(U) splitting}\ee
where $\Omega^\bt(U)$ is the algebra of differential forms on $U$, $\C$ is some fixed finite dimensional unital dg Frobenius algebra with differential $D$, product $m$ and pairing $\Pi$ of degree $-n_\C$ (with $\Tr$ the corresponding trace). We denote the unit of $\C$ by $\1$ and impose that
$\C$ is equipped with the splitting
\be \C=\RR\cdot\1\oplus\bar\C \label{C splitting}~,\ee
where $\bar\C$ is an ideal. The ring $\Omega^\bt(U)\otimes\C=C^\infty(U)\oplus \ldots$ inherits the splitting and the restriction homomorphisms must respect the splitting.  Moreover we require the existence of a morphism of sheaves of complexes
$$\Tr_J : {\cal J} \rightarrow \Omega^\bullet~,$$
 where $\Omega^\bullet$ is sheaf of differential forms. This morphism has a degree $-n_\C$ and locally has the form $\id_U \otimes \Tr$.
  Therefore the set of the global sections of $\J$ is equipped with the structure of dg Frobenius algebra with the trace given by the 
   composition of $\Tr_J$  and integration of
differential forms over $\N_0$ and with total differential $d_{\N_0}+D$.

\textbf{The target:} A $\ZZ$-graded manifold $\M$ with body $\M_0$, equipped with symplectic form $\omega\in\Omega^2(\M)$ of degree $n$ and a function $\Theta\in C^\infty(\M)$ of degree $n+1$, satisfying $\{\Theta,\Theta\}=0$. Here $n:=\dim \N_0+n_\C-1$.

We define the space of BV fields $\FF=\Maps(\J,\M)$ as the space of morphisms of ringed spaces from $(\N_0,\J)$ to $(\M_0,\M)$.  Let $(\phi,\Phi)\in\FF$ and let $U\subset\M_0$ be a coordinate neighborhood such that $\M(U)\sim C^\infty(U)\otimes S(V^*)$. Let $\{e_\zeta\}$ be some basis in $\C$ with $e_0=\1$ and $e_\zeta\in\bar\C$ for $\zeta\not=0$; let $\{X^A\}=\{x^\mu;\xi^m\}$ be local coordinates on $\M$ (i.e. $\{x^\mu\}$ are coordinates on $U\subset \M_0$ and $\{\xi^m\}$ are coordinates on $V$), and let $\{u^a\}$ be local coordinates on $\N_0$. Then the BV field $(\phi,\Phi)\in\FF$ is locally described by the {\it superfields} that are the values of $\Phi^*(U)$ on the generators $\{X^A\}$ of $\M(U)$:
\be \Phi^*(U): X^A\mapsto \Phi^{A\zeta}_0(u)e_\zeta+\Phi^{A\zeta}_a(u)\theta^a e_\zeta+\Phi^{A\zeta}_{a_1 a_2}(u)\theta^{a_1}\theta^{a_2}e_\zeta+\cdots
=\Phi^{A\zeta}(u,\theta)e_\zeta ~.\label{sheaf superfield}\ee
Here $\theta^a:=d_{\N_0}u^a$ are the odd generators of $\Omega^\bt(\phi^{-1}(U))$.
The splitting condition (\ref{C splitting}) allows to define the coefficients of $1$ that define a local map $\phi_U:\phi^{-1}(U)\rightarrow U$ as $x^\mu\circ\phi_U(u)=\Phi^{\mu 0}_0(u)$. By using a standard argument  we see that $\phi_U(u)=\phi(u)$, {\it i.e.} we recover the global map $\phi$.

Let us denote the symplectic form as $\omega=dX^A\omega_{AB}dX^B$; we construct the degree -1 symplectic form $\Omega$ on $\FF$ as
\begin{multline} 
\Omega=\int\limits_{\N_0}\Tr\left(\delta\Phi^A\,\omega_{AB}\,\delta\Phi^B\right)=\\
=(-1)^{(|A|+\dim\N_0+1)\cdot |\zeta|+(n+1)n_\C} \int\limits_{\N_0}\delta\Phi^{A\zeta}\;\Pi(e_\zeta,e_\eta)\;\omega_{AB}\;\delta\Phi^{B\eta} ~,
\end{multline}
where $\Phi^A=\Phi^*(X^A)$ is the right hand side of (\ref{sheaf superfield}) and the expression under trace in the first line implicitly uses the multiplication $m$ in $\C$.
The abstract AKSZ action is
\begin{equation}
S[\Phi]=
\underbrace{\frac{1}{2}\int\limits_{\N_0}\Tr\left(\Phi^A\omega_{AB}(d_{\N_0}+D)\Phi^B\right)}_{S_{kin}}+
\underbrace{(-1)^{n+1}\int\limits_{\N_0}\Tr(\Phi^*(\Theta))}_{S_{int}}~.
\end{equation}
The kinetic term may be also written as
\begin{multline}
S_{kin}=(-1)^{(|A|+\dim\N_0)\cdot |\zeta|+n_\C \dim \N_0}
\cdot\int\limits_{\N_0}\left((-1)^{|A|+\dim\N_0}\frac{1}{2}\Phi^{A\zeta}\;\Pi(e_\zeta,D e_\eta)\;\omega_{AB}\;\Phi^{B\eta}
+\right.
\\
\left.+(-1)^{|\zeta|}\frac{1}{2}\Phi^{A\zeta}\;\Pi(e_\zeta,e_\eta)\;\omega_{AB}\;d_{\N_0}\Phi^{B\eta}\right)~.
\end{multline}
If $\Theta$ is expanded in local coordinates on $\M$ as (\ref{Theta Taylor}), then the expansion analogous to (\ref{S_int Taylor}) in this setting is
$$S_{int}=(-1)^{n+1+n\cdot n_C}\int\limits_{\N_0}\Theta_A \Tr(e_\zeta)\;\Phi^{A\zeta}+(-1)^{(|A|+\dim\N_0+1)\cdot |\zeta|}\frac{1}{2}\Theta_{AB} \Tr(m(e_\zeta, e_\eta))\;\Phi^{A\zeta}\Phi^{B\eta}+\cdots$$

We again claim that $\{S,S\}=0$.

\begin{example}{\rm Taking the trivial dg Frobenius algebra $\C=\RR$, we obtain standard AKSZ models on $\Maps(T[1]\N_0,\M)$. }\end{example}

\begin{example}{\rm Taking $\N_0=pt$, specifying some finite dimensional  dg Frobenius algebra $\C$ 
with
splitting (\ref{C splitting}) and a finite-dimensional target $(\M,\omega,\Theta)$, we obtain a finite-dimensional abstract AKSZ model. A particular class of such sources is provided by the de Rham cohomology of connected closed orientable manifolds $\C=H^\bt_{dR}(\Sigma)$, viewed as dg Frobenius algebras with $D=0$ and $\Pi$ associated to Poincar\'e duality. Since $\Sigma$ is connected, splitting (\ref{C splitting}) is automatic: $\C=\RR\cdot\1\oplus H^{\geq 1}_{dR}(\Sigma)$. For these models $S_{kin}=0$, and the AKSZ action is just the pull-back of the function $\Theta$ on the target. These are the examples that we studied in (\ref{fin-dim-AKSZaction}) when $\Sigma$ is two-dimensional and in (\ref{reducedBV3d}) when $\Sigma$ is three-dimensional. }\end{example}

\begin{example}\label{ex_principal_bundle}(Source given by fiber cohomology of a fiber bundle.) {\rm 
%
Suppose $E$ is a fiber bundle over $\N_0$ with typical fiber $F$ (closed connected orientable manifold), endowed with a flat connection $\nabla_E$. Then we define fiber differential $d_\fib: \Omega^\bt(E)\ra\Omega^{\bt+1}(E)$ as follows: for 0-forms $f\in C^\infty(E)$ we define $d_\fib$ by the property $i_v(d_\fib f):=v_\perp(f)$ (where $i_v$ is the convolution with arbitrary vector field $v\in\mathrm{Vect}(E)$ and the projection to fiber $v\mapsto v_\perp$ is the projection to second term in the splitting of tangent bundle $TE=T_{||}E\oplus T_\perp E$ defined by the connection $\nabla_E$); then we extend $d_\fib$ to all forms on $E$ by Leibniz rule and property $d_E d_\fib+d_\fib d_E=0$ (where $d_E$ is the de Rham differential on $E$). Flatness of $\nabla_E$ implies that $d_\fib^2=0$. Then we construct sheaf $\J$ over $\N_0$ as the cohomology of $d_\fib$:
$$\Gamma(\N_0,\J)=H^\bt_{d_\fib}(\Omega^\bt(E))~.$$
Locally $\J$ splits as (\ref{J(U) splitting}) with $\C=H^\bt(F)$ (de Rham cohomology of the fiber).

In particular we can take a trivial fiber bundle $E=\N_0\times F$ with canonical flat connection $\nabla_E$. Then $d_\fib$ is just the de Rham differential along fiber $d_\fib=\id_{\N_0}\otimes d_F$
and the fiber cohomology sheaf $\J$ splits globally: $$\Gamma(\N_0,\J)=\Omega^\bt(\N_0)\otimes H^\bt(F)~.$$

Abstract AKSZ models with source $\J$ and some target $(\M,\omega,\Theta)$ may in some cases arise as a partial reduction of usual AKSZ models on $\Maps(T[1]E, \M)$. The simplest example here is: $E$ is the 2-torus, viewed as a trivial bundle over circle $\N_0={\mathbb S}^1$ with fiber a circle $F={\mathbb S}^1$, and  $\M$ is a Poisson manifold. The corresponding abstract AKSZ model 
 is a partial reduction of Poisson sigma model on torus.

We hope to explore this set of examples in more detail in a future publication.
}\end{example}

\begin{remark}{\rm Global sections $\Gamma(\N_0,\J)$ of the sheaf $\J$ themselves form a dg Frobenius algebra with differential $D_\J=d_{\N_0}+D$ (where $d_{\N_0}$ is the de Rham differential on $\N_0$), multiplication $m_\J$ coming from wedge product of forms and multiplication $m$ on $\C$, and with the pairing $\Pi_\J(\chi,\psi)=\int_{\N_0}\Tr (m(\chi,\psi))$. When the target is a graded vector space, we can apply the vector space construction of Subsection \ref{vector_space_version} with  $\Gamma(\N_0,\J)$ as source. We easily verify that the two constructions coincide.
Remark that requirement (\ref{C splitting}) is not needed in the vector space version, so that more general Frobenius algebras are allowed in the vector space version. } \end{remark}

\bigskip\bigskip

\section{Conclusions}
 \label{summary}

In this paper we studied a canonical reduction of the AKSZ-BV field theory to a finite dimensional BV theory which governs the semi-classical approximation.  As illustration of the general construction, we discussed the two dimensional Poisson sigma model and the three dimensional Courant sigma model.

Our main perspective has been the odd symplectic reduction of the infinite dimensional manifold of fields.  It is important to remark that one can look at the reduced action $S_{BV}$ as the leading contribution in the effective BV action which controls the low energy fields (we can call
them either "constant" maps or zero modes). It is convenient to consider the idea of effective BV theories suggested by Losev \cite{losev} (see \cite{Mnev:2006ch}, \cite{Costello:2007ei} and \cite{Cattaneo:2008yf} for further developments). In fact, given any embedding of the cohomology in the space of forms of the source, one can look at the reduced variables as "infrared" degrees of freedom of the full theory. The effective action is then defined by integrating over the "ultraviolet" degrees of freedom and, in the perturbative approach, is a series in $\hbar$ and in the hamiltonian function $\Theta$ of the target. From this point of view, the reduced $BV$ manifolds that we studied in this paper are the spaces of the infrared degrees of freedom and the action $S_{BV}$ is the lowest order in the expansion of the effective action. In principle one can calculate the corrections by applying Feynman diagrams techniques.

\medskip

From the examples considered in this paper, it is natural to consider the generalization of the AKSZ construction \cite{Alexandrov:1995kv}. In Section \ref{reduction_cohomology} we observed that the
reduced BV theory can be described in terms of "supermaps" to the target graded manifold. The novelty is that the formal variables of the source manifold have to satisfy some constraints and thus they cannot be anymore considered as the coordinates of a graded manifold. This is the generalization of the AKSZ construction that we introduced in Section \ref{generalizationAKSZ}. The examples that appeared in this paper have sources that are commutative graded algebras seen as sheaves over a point and so are zero dimensional TFT's. One can consider generalized AKSZ theories in any dimension, as we described for instance in the Example (\ref{ex_principal_bundle}), that will be the object of future study.

\medskip

Finally, one can consider more general type of BV reductions, not necessary
to the constant map configurations. Moreover many ideas presented here can be applied to a wider setup than simply AKSZ-BV theories. For example, it could be interesting to study the reduction of the two dimensional BV theories described in \cite{Zucchini:2004ta, Zucchini:2008hn}.

\bigskip\bigskip

\noindent{\bf\Large Acknowledgement}:
\bigskip

\noindent We thank  Alberto Cattaneo, Jian Qiu, Dmitry Roytenberg and Gabriele Vezzosi
  for the discussions. 
P.M. also thanks Nikolai Mn\"ev and Nikolai Reshetikhin for discussions.
We are  happy to
thank the program ``Geometrical Aspects of String Theory'' at Nordita,
where part of this work was carried out.
 M.Z. thanks INFN Sezione di Firenze  and Universit\`a di Firenze
  where part of this work was carried out.
 The research of M.Z. was
supported by VR-grant 621-2004-3177 and by VR-grant 621-2008-4273.

\bigskip\bigskip

\appendix

\section{Computation of Berezinian (\ref{berezinian3d})}
\label{ap-ber}

We show here that the volume form introduced in (\ref{berezinian3d}) is globally defined on ${\rm Map}(X_\Sigma,\M)$. The coordinates $z=({\mathsf X}^\mu,{\boldsymbol \xi}_a,{\boldsymbol P}_\mu)$ defined as coefficients of the superfields (\ref{reducedsuperfields3d}) depend on the choice of coordinates $\{x^\mu\}$ on $M$ and of a trivialization $\{e_a\}$ of $E$. If we change to coordinates $\{y^i=y^i(x)\}$ on $M$ and to trivialization $\{e_\alpha=t_{\alpha}^a(x) e_a\}$, the coordinates on ${\rm Map}(X_\Sigma,\M)$ change to $\tilde{z}=({\mathsf X}^i,{\boldsymbol \xi}_\alpha,{\boldsymbol P}_i)$ accordingly as
\begin{equation}\label{reducedtransf3d}
{\mathsf X}^i=y^i({\mathsf X})\;,~~~ {\boldsymbol \xi}^\alpha = t^\alpha_a({\mathsf X}){\boldsymbol \xi}^a ~,~~~~ {\boldsymbol P}_i =\frac{\partial x^\mu}{\partial y^i}({\mathsf X}){\boldsymbol P}_\mu  + \frac{1}{2} {\boldsymbol \xi}^a{\boldsymbol \xi}^b \frac{\partial t_a^{\alpha}}{\partial y^i}({\mathsf X})g_{\alpha\beta}t^{\beta}_b({\mathsf X})\;.
\end{equation}
The quadratic term in the transformation of ${\boldsymbol P}_i$ can be removed by introducing a connection on the vector bundle $E$. In fact, the coordinate ${\boldsymbol P}_\mu^\Gamma = {\boldsymbol P}_\mu + \frac{1}{2}\Gamma_{\mu a}^b {\boldsymbol \xi}^ag_{bc}{\boldsymbol \xi^c}$ transforms as a tensor
$$
{\boldsymbol P}_i^\Gamma =\frac{\partial x^\mu}{\partial y^i}({\mathsf X}){\boldsymbol P}_\mu^\Gamma \;.
$$
It can be easily checked that the Berezinian of the transformation from $\{{\mathsf X}^\mu,{\boldsymbol \xi}^a,{\boldsymbol P}_\mu\}$ to $\{{\mathsf X}^\mu,{\boldsymbol \xi}^a,{\boldsymbol P}_\mu^\Gamma\}$ is one so that the coordinate volume forms are the same. Equivalently, in order to compute the Berezinian of the transformation $\tilde{z}(z)$ we are allowed to ignore the quadratic term in the transformation of ${\boldsymbol P}$ in (\ref{reducedtransf3d}).

The final result is that the berezinian of the transformation matrix $I=(\frac{\partial \tilde{z}}{\partial z})$ is
$${\rm Ber} I = {\rm Ber} \left(\begin{array}{cc} I_{00} & I_{01} \cr I_{10} &I_{11}\end{array}\right) = (\det t^{\alpha}_a)^{2(1-b_1)}\;.$$
We first compute it with respect to the transformation $y^i=y^i(x)$ with fixed trivialization. Let us order the relevant coordinates $z=\{z_0,z_1\}$, where $z_0$ are the even ones and $z_1$ are the odd ones, as $z_0=(x^\mu,\gamma_\mu,F_{\mu I},\alpha^{+\mu}_I)$ and $z_1=(\alpha^I_\mu,F^{+\mu I},\gamma^{+\mu},x^+_\mu)$. It is important to see where the zeros are located in $I$, so that at the end only few matrix elements (the diagonal ones) enter the result. By inspection of the degree in (\ref{reducedtransf3d}) we can easily write
\begin{eqnarray*}
I_{00}=\frac{\partial\tilde{z}_0}{\partial z_0} = \left(\begin{array}{cc} \begin{array}{cc}\frac{\partial y^i}{\partial x^\mu }&0\cr * & \frac{\partial \gamma_i}{\partial \gamma_\mu } \end{array}& \begin{array}{cc}0&0\cr 0& 0\end{array}\cr * & \begin{array}{cc} \frac{\partial F_{iI}}{\partial F_{\mu J} }& * \cr 0 & \frac{\partial \alpha^{+i}_I}{\partial \alpha^{+\mu}_J }\end{array} \end{array}\right) &
I_{01}=\frac{\partial\tilde{z}_0}{\partial z_1} = \left(\begin{array}{cc} \begin{array}{cc}0&0\cr 0& 0\end{array} & \begin{array}{cc}0&0\cr 0& 0\end{array}\cr * &  \begin{array}{cc}0&0\cr 0& 0\end{array}\end{array}\right) \cr
I_{11}=\frac{\partial\tilde{z}_1}{\partial z_1} = \left(\begin{array}{cc} \begin{array}{cc}\frac{\partial \alpha_i^I}{\partial \alpha_\mu^J }& * \cr 0 & \frac{\partial F^{+iI}}{\partial F^{+\mu J} } \end{array}& \begin{array}{cc}0&0\cr 0& 0\end{array}\cr * & \begin{array}{cc} \frac{\partial \gamma^{+i}}{\partial \gamma^{+\mu} }& 0 \cr * & \frac{\partial x_{+i}}{\partial x_{+\mu}}\end{array} \end{array}\right) &
I_{10}=\frac{\partial\tilde{z}_1}{\partial z_0} = \left(\begin{array}{cc} *& \begin{array}{cc}0&0\cr 0& 0\end{array}\cr * & * \end{array}\right)\;,
\end{eqnarray*}
where the block structure is easily understood. It is then easy to compute that
$$
I_{01} I_{11}^{-1} I_{10} = \left(\begin{array}{cc}\begin{array}{cc}0&0\cr 0& 0\end{array} &\begin{array}{cc}0&0\cr 0& 0\end{array}\cr * & \begin{array}{cc}0&0\cr 0& 0\end{array}\end{array}\right)
$$
so that
$$
{\rm Ber} I = \det (I_{00}- I_{01} I_{11}^{-1} I_{10})/\det I_{11} = \det I_{00}/\det I_{11} = 1\;.
$$
Consider now the change of trivialization $e_\alpha = t_{\alpha}^a(x) e_a$, without changing coordinates $\{x^\mu\}$. Let us order the relevant coordinates as follows: the even ones are $z_0=(x^\mu,\alpha^{+\mu}_I,A^{aI},\beta^+_a)$ and the odd ones are $z_1=(F^{+\mu I},\gamma^{+\mu},\beta^a,A^+_{aI})$. Since we are ignoring the quadratic terms in (\ref{reducedtransf3d}) the coefficients of ${\boldsymbol P}_\mu$ do not appear. Then we compute
\begin{eqnarray*}
I_{00}= \left( \begin{array}{cc} \begin{array}{cc}\delta^\mu_\nu & 0 \cr 0 &\delta^I_J\delta^\mu_\nu\end{array} & \begin{array}{cc}0&0\cr 0& 0\end{array}\cr * & \begin{array}{cc} \frac{\partial A^{\alpha I}}{\partial A^{a J} }& 0 \cr * & \frac{\partial \beta^+_\alpha}{\partial \beta^+_a}\end{array} \end{array}\right) &
I_{01} = \left(\begin{array}{cc} \begin{array}{cc}0&0\cr 0& 0\end{array} & \begin{array}{cc}0&0\cr 0& 0\end{array}\cr * & \begin{array}{cc}*&0 \cr * & *\end{array} \end{array}\right) \cr
I_{11}= \left(\begin{array}{cc} \begin{array}{cc}\delta^\mu_\nu\delta^I_J&0\cr 0 & \delta_\mu^\nu \end{array}& \begin{array}{cc}0&0\cr 0& 0\end{array}\cr * & \begin{array}{cc} \frac{\partial \beta^\alpha}{\partial \beta^a }& 0 \cr * & \frac{\partial A^+_{\alpha I}}{\partial A^+_{a I}}\end{array} \end{array}\right) &I_{10}= \left(\begin{array}{cc} \begin{array}{cc}0&0\cr 0& 0\end{array}& \begin{array}{cc}0&0\cr 0& 0\end{array}\cr * & \begin{array}{cc} 0 & 0 \cr * & 0\end{array} \end{array}\right)\;.
\end{eqnarray*}
We then compute that
$$
I_{01} I_{11}^{-1} I_{10} = \left(\begin{array}{cc} \begin{array}{cc}0&0\cr 0& 0\end{array}& \begin{array}{cc}0&0\cr 0& 0\end{array}\cr * & \begin{array}{cc}0& 0\cr * & 0\end{array}\end{array}\right)
$$
 and finally we get
$$
{\rm Ber} I = \det (I_{00}- I_{01} I_{11}^{-1} I_{10})/\det I_{11} = \det I_{00}/\det I_{11} = (\det t^\alpha_a)^{2(b_1 -1)} \;.
$$

\Section{Relative (co)homology}
\label{relative_homology}

We recall in this appendix basic facts about relative (co)homology and Lefschetz duality for manifolds with boundary \cite{Hatcher}.

Let $\Sigma$ be a smooth manifold of dimension $d$ with boundary $\partial\Sigma$. The relative $k$-chains with real coefficients are defined as $C_k(\Sigma,\partial\Sigma)=C_k(\Sigma)/C_k(\partial\Sigma)$, {\it i.e.} the chains in $\Sigma$ modulo the chains in $\partial\Sigma$. We will always work with real coefficient and we will omit it in the notation. The usual boundary $\partial$ goes to the quotient and defines the relative homology $H_k(\Sigma,\partial\Sigma)$. An alternative description of chains is obtained by defining $C_k'(\Sigma,\partial\Sigma)=\{(c_k,\sigma_{k-1})\ ,\ c_k\in\ C_k(\Sigma), \sigma_{k-1}\in C_{k-1}(\partial\Sigma)\}$ with boundary $\partial(c_k,\sigma_{k-1})=(\partial c_k + (-)^k \sigma_{k-1},\partial\sigma_{k-1})$. It is easy to check that the map $(c_k,\sigma_{k-1})\in C_k'(\Sigma,\partial\Sigma)\rightarrow \underline{c_k}\in C_k(\Sigma,\partial\Sigma)$ is a quasisomorphism.

The exact sequence $0\rightarrow C_k(\partial\Sigma)\rightarrow C_k(\Sigma)\rightarrow C_k(\Sigma,\partial\Sigma)\rightarrow 0$ gives rise to the long exact sequence in homology
\begin{equation}\label{exact_sequence_homology}
\ldots\rightarrow H_k(\partial\Sigma) \rightarrow H_k(\Sigma) \rightarrow H_k(\Sigma,\partial\Sigma) \rightarrow H_{k-1}(\partial\Sigma)\rightarrow \ldots  \;,
\end{equation}
where the last map sends $[\underline{c}]\in H_k(\Sigma,\partial\Sigma)\rightarrow [\partial c]\in H_{k-1}(\partial\Sigma)$, for some $c\in C_k(\Sigma)$.

The complex of relative cochains $C^\bullet(\Sigma,\partial\Sigma)$ can be described as the restriction of de Rham complex to those forms $\omega$ whose restriction $\omega|_{\partial\Sigma}$ to the boundary is zero. We denote the relative cohomology as $H_{dR}(\Sigma,\partial\Sigma)$. By the universal coefficient theorem we have that $H_{dR}^\bullet(\Sigma)=H(\Sigma)^*_\bullet$. The alternative description for $k$-relative cochains is $C'{}^k(\Sigma,\partial\Sigma)=\Omega^k\Sigma\oplus\Omega^{k-1}\partial\Sigma$, with differential $d(\omega_k,\nu_{k-1})=(d\omega_k, d\nu_{k-1}- (-)^k \omega_k|_{\partial\Sigma})$. The map $\omega_k\in C^k(\Sigma,\partial\Sigma)\rightarrow (\omega_k,0)\in C'{}^k(\Sigma,\partial\Sigma)$ is a quasisomorphism. The pairing is then defined as
$$
\langle\omega_k, \underline{c_k}\rangle = \int_{c_k} \omega_k  \;.
$$
or alternatively as
$$
\langle (\omega_k,\nu_{k-1}),(c_k,\sigma_{k-1})\rangle = \int_{c_k} \omega_k + \int_{\sigma_{k-1}} \nu_{k-1}\;.
$$

The above notion of relative (co)homology makes sense for any subspace of $\Sigma$, in particular we can consider (union of) components of $\partial\Sigma$. If $\partial\Sigma=\cup_{i\in I} \partial_i\Sigma$ and $\partial_J\Sigma=\cup_{i\in J}\partial_i\Sigma$, for $J\subset I$, we will consider the relative homology $H(\Sigma,\partial_J\Sigma)$ and cohomology $H_{dR}(\Sigma,\partial_J\Sigma)$. Let $I=I_1\cup I_2$, with $I_1\cap I_2=\emptyset$; the choice of the fundamental class $[\Sigma]\in H_d(\Sigma,\partial\Sigma)$ determines the following isomorphism (for a proof see Theorem 3.43 of \cite{Hatcher})
$$
H_k(\Sigma, \partial_{I_1}\Sigma)= H_{dR}^{d-k}(\Sigma,\partial_{I_2}\Sigma)\;.
$$
In particular, the case $I_1=\emptyset$ or $I_2=\emptyset$ is known as {\it Lefschetz duality},
$$
H^k_{dR}(\Sigma,\partial\Sigma) \sim H_{d-k}(\Sigma)\;,~~~~   H^k_{dR}(\Sigma)\sim H_{d-k}(\Sigma,\partial\Sigma)\;.
$$

Let us describe more explicitly the case $d=2$ and consider a compact surface $\Sigma_{g,n}$ of genus $g$ and $n$ boundary components.

Since $H_0(\Sigma_{g,n},\partial\Sigma_{g,n})=H_2(\Sigma_{g,n})=0$, by Lefschetz duality we get that $H^2_{dR}(\Sigma_{g,n})=H^0_{dR}(\Sigma_{g,n},\partial\Sigma_{g,n})=0$. The Lefschetz duality in degree one can be seen as the non degeneracy of the pairing $H^1_{dR}(\Sigma_{g,n})\otimes H^1_{dR}(\Sigma_{g,n},\partial\Sigma_{g,n})\rightarrow \R$, $(a,b)\rightarrow \int_{\Sigma_{g,n}}a\wedge b$. Equivalently, for any basis $\{c_I\}$ for $H^1_{dR}(\Sigma_{g,n})$ and $\{g_{I'}\}$ for $H_{dR}^1(\Sigma_{g,n},\partial \Sigma_{g,n})$ the matrix $A^{(g,n)}_{II'}=\int_{\Sigma_{g,n}} c_I\wedge g_{I'}$ is non degenerate. Let us denote with $A^{(g,n)}{}^{I'J}$ the inverse matrix. It will be useful even the (possibly degenerate) matrix $B^{(g,n)}_{I'J'}=\int_{\Sigma_{g,n}}g_{I'}\wedge g_{J'}$.


\end{document}